\documentclass[twocolumn]{IEEEtran}

\ifCLASSINFOpdf
\else
\fi
\usepackage{dblfloatfix}    

\usepackage{cite}
\usepackage{amsmath,amssymb,amsfonts}
\usepackage{bbm}

\usepackage{graphicx}
\usepackage{textcomp}
\usepackage{xcolor}
\usepackage{colortbl} 

\usepackage{url} 

\usepackage{bm}
\usepackage{supertabular}
\usepackage{amsthm}
\theoremstyle{definition}

\newtheorem{remark}{Remark}
\usepackage[linesnumbered,ruled,vlined]{algorithm2e}
\usepackage{algpseudocode}
\algrenewcommand{\algorithmiccomment}[1]{\hfill// #1}

\usepackage{pbox}
\usepackage{float} 
\usepackage{subcaption} 

\usepackage{boldline,multirow}
\usepackage{array}

\usepackage{pbox}

\usepackage{dblfloatfix}

\usepackage[normalem]{ulem}

\usepackage{cancel} 

\newcommand*\diff{\mathop{}\!\mathrm{d}}

\definecolor{anti-flashwhite}{rgb}{0.95, 0.95, 0.96}

\begin{document}
	
	\setlength{\abovedisplayskip}{2pt}
	\setlength{\belowdisplayskip}{2pt}

	\title{Handoffs in User-Centric Cell-Free MIMO Networks: A POMDP Framework}
	%
	%
	\author{Hussein~A.~Ammar\IEEEauthorrefmark{1},~\IEEEmembership{Graduate Student Member,~IEEE}, 
		Raviraj~Adve\IEEEauthorrefmark{1},~\IEEEmembership{Fellow,~IEEE},
		Shahram~Shahbazpanahi\IEEEauthorrefmark{2}\IEEEauthorrefmark{1},~\IEEEmembership{Senior Member,~IEEE},
		Gary~Boudreau\IEEEauthorrefmark{3},~\IEEEmembership{Senior Member,~IEEE}, 
		and~Kothapalli~Venkata~Srinivas\IEEEauthorrefmark{3},~\IEEEmembership{Member,~IEEE}
		\thanks{This work was supported in part by the Natural Sciences and Engineering Research Council (NSERC) of Canada and in part by Ericsson Canada. An earlier version of this paper was presented in part at the 2022 IEEE Global Communications Conference (GLOBECOM)~\cite{conf_POMDP_10001033}.}
		\thanks{
			\IEEEauthorrefmark{1}H. A. Ammar and R. Adve are with the Edward S. Rogers Sr. Department of Electrical and Computer Engineering, University of Toronto, Toronto, ON M5S 3G4, Canada (e-mail: ammarhus@ece.utoronto.ca; rsadve@comm.utoronto.ca).
		}
		\thanks{
			\IEEEauthorrefmark{2}S. Shahbazpanahi is with the Department of Electrical, Computer, and Software Engineering, Ontario Tech University, Oshawa, ON L1H 7K4, Canada. He also holds a Status-Only position with the Edward S. Rogers Sr. Department of Electrical and Computer Engineering, University of Toronto.
		}
		\thanks{
			\IEEEauthorrefmark{3}G. Boudreau is with Ericsson Canada, Ottawa, ON K2K 2V6, Canada. K. V. Srinivas was with Ericsson Canada, he is now with with Motorola Mobility/Lenovo, Ottawa, ON K1Z 8R1, Canada.
		}
	}
	
	
	%
	%
	
	\maketitle 

	
	\newcommand*{\nOfHOsTtrigLower}{$47\%$}
	\newcommand*{\nOfHOsThreshtrig}{$70\%$}
	\newcommand*{\CDFTtrigLower}{$82\%$}
	\newcommand*{\CDFThreshtrig}{$90\%$}

	\begin{abstract}
		We study the problem of managing handoffs (HOs) in user-centric cell-free massive MIMO (UC-mMIMO) networks. Motivated by the importance of controlling the number of HOs and by the correlation between efficient HO decisions and the temporal evolution of the channel conditions, we formulate a partially observable Markov decision process (POMDP) with the state space representing the discrete versions of the large-scale fading and the action space representing the association decisions of the user with the access points (APs). We develop a novel algorithm that employs this model to derive a HO policy for a mobile user based on current and future rewards. To alleviate the high complexity of our POMDP, we follow a divide-and-conquer approach by breaking down the POMDP formulation into sub-problems, each solved separately. Then, the policy and the candidate pool of APs for the sub-problem that produced the best total expected reward are used to perform HOs within a specific time horizon. We then introduce modifications to our algorithm to decrease the number of HOs. The results show that half of the number of HOs in the UC-mMIMO networks can be eliminated. Namely, our novel solution can control the number of HOs while maintaining a rate guarantee, where a~\nOfHOsTtrigLower-\nOfHOsThreshtrig\ reduction of the cumulative number of HOs is observed in networks with a density of~$125$ APs per~${\rm km}^2$. Most importantly, our results show that a POMDP-based HO scheme is promising to control HOs.
	\end{abstract}
	
	%
	\IEEEpeerreviewmaketitle
	
	%
	%
	%
	\section{Introduction}
	User-centric cell-free massive multiple-input multiple-output (UC-mMIMO) networks provide interesting capabilities to improve both the spectral and energy efficiencies~\cite{cellFreeVersusSmallCells7827017, 8845768, 8097026}. This network paradigm provides a relatively uniform spatial coverage across the network area~\cite{8449213, interdonato2019ubiquitous} by eliminating the cell-based connections for users. 
	In cell-free MIMO networks, users form individual serving sets comprising nearby access points (APs) (or radio units)~\cite{ammar9519163, ammar9570126}.
	
	
	As indicated in the survey paper in~\cite{ammar9650567}, the UC-mMIMO scheme is a promising candidate for future wireless network architectures. However, as noted there, this scheme coupled with a possible dense deployment of APs can negatively affect the performance of users having high-speed mobility profiles. The main issue is that a mobile user may experience unstable connections due to frequent handoffs (HOs). In conventional cellular networks, a HO or handover transfers a communication session from one channel to another, which mostly involves changing the \emph{single} serving base station (BS). In a cell-free network, given the multiple connections between the user and the network, HOs could be far more frequent. Herein, we refer to HO as the process of adding/dropping a AP to/from the serving set of the user.	
	
	The problem of frequent HOs in UC-mMIMO networks is an interesting topic that has been rarely visited. The authors in~\cite{mobilityMMwaveCellFree9616361} partially consider minimizing the number of HOs in this network scheme with millimeter wave communication. However, the approach does not focus on HOs. Still this study reinforces our expectation that HO could be a serious problem for UC-mMIMO networks. In this regard,~\cite{mobilityMMwaveCellFree9616361} mentions that frequent HOs can negatively affect performance. HOs in cell-free networks are more challenging than those in a conventional cellular network, and instead of proposing simple HO decisions, we now need to set a user-APs \textit{association strategy}~\cite{zaher2022soft_10316237}. 
	Specifically, limiting the unnecessary number of changes in the set of serving APs is an important goal. Overall, HO strategies to exploit the benefits provided by cell-free communications are crucial~\cite{beerten2023cell}. Indeed, a HO \textit{strategy}, as opposed to simple HO \textit{decisions}, is an advanced framework that can provide stability of connections while taking into account \textit{long-term} performance.

	Overall, the distinctions in HOs within the UC-mMIMO network scheme, as opposed to conventional networks, can be summerized in a few key points. First, HO decisions are executed for a serving set of multiple APs without a cellular border~\cite{PDPUsercentricVsDisjoint8969384}, as opposed to a single serving BS covering a predefined cellular area in a conventional network. This necessitates increased control plane traffic. It also complicates the trigger for HOs in UC-mMIMO networks because the serving set lacks clear borders as opposed to cells. Consequently, the UC-mMIMO network offers a high degree of flexibility in defining the HO triggers based on the network performance and behavior objectives. Second, when contrasting with conventional networks, it becomes evident that the handling of numerous connections to APs in UC-mMIMO networks necessitates the implementation of active control solutions for achieving seamless performance~\cite{10264861}. This highlights the importance of our study.

	
	The literature shows that user-mobility imposes challenges not only on HOs but also on channel estimation. The study in~\cite{CSIAgingzheng2020cell} derives the achievable rate for the uplink of cell-free massive MIMO network when using large-scale fading (LSF) decoding and a matched-filter receiver. Nonetheless, the study does not address the problem of HOs, instead, it focuses on \textit{channel aging}~\cite{CSIAging6608213}, which is the mismatch between the channel state information (CSI) at the time the channel was estimated and at the time it is used for data transmission.
	
	The study in~\cite{9416909} investigates the impact of channel aging on the performance of cell-free massive MIMO networks and derives closed-form expressions for the achievable rate in the uplink and downlink using coherent and noncoherent transmissions~\cite{9500429}. 
	Moreover, the authors in~\cite{ChAgingPhaseNoise9471851} consider user-mobility while focusing on phase noise. The phase noise is used in the analysis due to an assumption of imperfect local oscillators in the distributed access points, which could possibly be solved with accurate synchronization protocols. 
	Channel aging is also analyzed in the literature of massive MIMO networks, where predicting the channel using a finite impulse response (FIR) Wiener predictor~\cite{CSIAging6608213} and optimizing the duration of the communication frame~\cite{channelAgingMassiveMIMO8122014} are proposed to enhance performance.
	

	Due to the user-centric clustering, users may experience frequent HOs as they move, thus, a procedure to control the number of HOs is indeed favorable in UC-mMIMO networks. Provided this, we first analyze the effect of user-mobility on channel estimation and achievable data rate in the presence of channel aging, which places some limitations on the pilot scheme used in the pilot training phase~\cite{9416909}. Equipped with a closed-form expression for the spectral efficiency and the temporal correlation for the LSF experienced by the mobile user, we model the problem of HOs using a partially observable Markov decision process (POMDP).
	
	A Markov decision process (MDP) is a Markov process with feedback control implemented through a decision maker, while a POMDP is a generalization of an MDP in which the states of the process are not fully observable. In a POMDP, the decision maker derives a policy that decides the actions to be taken based on the noisy observations of the states. The history of observations and actions is used to construct a posterior distribution for the states which is commonly known as the beliefs~\cite{krishnamurthy2016partially, 9626007}. Importantly, algorithms to solve POMDPs have been well-investigated and developed~\cite{shani2013survey, lovejoy1991survey}.

	One popular choice for solving POMDPs/MDPs is a family of deterministic algorithms known as Dynamic Programming~\cite{smallWoodSondik, sondik1978optimal, 9626007}. These algorithms are categorized into value iteration and policy iteration methods. The value iteration algorithm involves updating the expected sum of discounted rewards for each state (or belief state in POMDPs), which is called the value function. 
	The policy iteration is a second standard algorithm to solve POMDPs/MDPs and it involves alternating between improving a policy and evaluating its performance. In this approach, the policy is iteratively refined until an optimal policy is found. Many algorithms for these categories have been developed and refined multiple times so that they can solve harder problems. An example of such a value iteration algorithm is the point-based value iteration (PBVI)~\cite{pineau2003point} which approximates a value iteration solution by selecting a small set of representative belief points and then tracking the value and its derivative for those points only. This approach allows us to solve problems with large sizes, and it is adopted in our paper.
	
	Reinforcement learning (RL) presents an alternative approach to learn a policy that maps observations of an environment to actions. Unlike dynamic programming, RL does not require full knowledge of the MDP's dynamics, e.g., the transition probabilities and the reward function. Using RL, an agent can learn a policy through trial and error, exploring the environment and receiving rewards or penalties based on the actions taken; hence RL is an online learning approach. The policy obtained from RL can be a tabular policy when using algorithms such as Q-learning~\cite{watkins1992q}, or it can be a trained neural network when using deep RL (DRL) algorithms.
	
	Other methods to obtain policies for POMDPs/MDPs involve the use of machine learning frameworks such as the multi-armed bandits~\cite{slivkins2019introduction}, which can be thought as a special case of RL. In multi-armed bandits, the next state of the MDP does not depend on the action chosen by the agent.

	Unfortunately, all of these techniques suffer from exponential complexity in the number of states. Hence, the novelty in our study is two-fold: in modeling the problem of HO in UC-mMIMO networks as a POMDP and in proposing crucial procedures to control the computational complexity of solving this POMDP.
	
	Our usage of POMDP has many aspects. First, the HO problem depends on subsequent events occurring over time, for example, we may not want to initiate HO very frequently, or we may wish to consider the predicted status of the system in the future when performing HOs. 
	In this regard, a POMDP is used to derive a HO policy that takes into account future rewards within the time horizon of the model. Second, the partial observability feature of POMDP allows to take informed decisions for HOs even when we cannot observe \emph{all} the channels between the user and the APs. 
	Third, the feedback control inherent in a POMDP provides an active control scheme for HOs based on the current observed state of the system.
	
	
	We set the state space of our POMDP model as the different combinations of the discrete versions for the LSF of the channels between the user and the APs. We set the action space as the connection decisions between the user and the APs. Hence, the policy derived by the POMDP determines the AP-association of the user when moving, thus producing a HO management platform for the user. To construct a complete and scalable solution for HOs, we develop an algorithm that controls the computational complexity as the network size increases. This is achieved through a divide-and-conquer approach. In this context, we select different pools of APs to construct POMDP sub-problems that are solved independently. We select the candidate POMDP sub-problem that produces the largest total expected reward. Then, the policy and the pool of APs for the selected POMDP sub-problem is used to determine the HO decisions for the user within the selected time horizon.
	
	To control the number of HOs, we modify our algorithm to allow us to decrease the number of HOs while maintainting a specific quality of service metric represented through a data rate threshold. An interesting feature of our POMDP-based HO is that, unlike LSF-based association, it can be used with partial knowledge of the LSF of the channels, also known as partial observability. The contributions of our paper can be summarized as follows:
	%
	%
	%
	%
	
	\label{page:varDimen}\newcommand{\varDimen}{We construct an algorithm that derives an updated HO policy with a manageable computational complexity even when the network size increases. This is done through a divide-and-conquer approach. 
	We emphasize that managing the computational complexity is necessary because POMDP problems suffer from the curse of dimensionality~\cite{madani1999undecidability}.}
	
	\begin{itemize}
		\item We develop the first study that employs a POMDP to model and control HOs for UC-mMIMO networks. In this regard, we take into account realistic conditions imposed by mobility such as channel aging and the evolution of LSF as a Markov chain. The benefits from our approach are numerous. The first benefit is providing the ability to account for hidden system states when studying HOs which allows for realistic assumptions compared to an MDP. The second benefit is deriving a HO \textit{policy} to manage HOs based on future rewards as the user moves. The third is presenting a robust and formal model for future studies of HOs using deep reinforcement learning (DRL) tools.
		\item \varDimen
		\item We introduce a control scheme to our algorithm which decreases the number of HOs or the HO rate for the user while maintaining a needed quality of service threshold. In this regard, our results show that our approach successfully decreases the number of HOs with a robust performance. Our results show a \nOfHOsTtrigLower-\nOfHOsThreshtrig\ reduction in the cumulative number of HOs in a network with density of $125$ APs per ${\rm km}^2$ compared to time-triggered and data rate threshold-triggered LSF-based HO approaches.
	\end{itemize}	
	
	The rest of the paper is organized as follows. Section~\ref{section:Model} presents the system model and the essentials needed to account for the effect of user-mobility. Section~\ref{sec:POMDP} formulates the management of HOs as a POMDP. Section~\ref{sec:Analysis} defines important steps to solve the POMDP, such as the POMDP sub-problems; this section also develops the algorithm used to control HOs. Section~\ref{sec:DecreasingNofHO} proposes modifications to the algorithm to decrease the number of HOs. Section~\ref{sec:Results} reports on our numerical results and findings. Finally, Section~\ref{section:Conclusion} concludes the paper.

	\emph{Notation:} Both lower and upper case letters (e.g., $a$ and $A$) represent scalars, while their bold counterparts ${\bf a}$ and ${\bf A}$ represent vectors and matrices, respectively. Tilde $\tilde{(\cdot)}$ over a letter represents a specific value taken by a random variable. 
	Operators $(\cdot)^{-1}$, $(\cdot)^T$, $(\cdot)^*$, and $(\cdot)^H$ denote the inverse, transpose, conjugate, and conjugate transpose, respectively. $\mathbb{E}\{\cdot\}$ represents statistical expectation, $\|\cdot\|$ and $|\cdot|$ are the vector and scalar Euclidean norms, 
	${\bf a} = {\rm diag} \left({\bf A}\right)$ selects the diagonal elements of ${\bf A}$ to construct a vector ${\bf a}$,
	and ${\bf I}_m$ is $m \times m$ identity matrix. For a set $\mathcal{A}$, $|\mathcal{A}|$ denotes its cardinality. Finally, $\mathbb{B}$, $\mathbb{R}$, $\mathbb{C}^{m\times 1}$, and $\mathbb{C}^{m\times n}$ represent binary numbers, real numbers, complex number $m\times 1$ vectors, and complex number $m\times n$ matrices, respectively.
	
	
	\vspace{-1em}
	\section{System Model}\label{section:Model}
	\vspace{-0.5em}
	We consider a network of $B$ APs (or access points) represented through the set $\mathcal{B}$. Each AP $b \in \mathcal{B}$ is equipped with $M$ antennas and serves single-antenna users. The users in the network, represented through the set $\mathcal{U}$, are served through the UC-mMIMO scheme. For each user $u \in \mathcal{U}$, a serving set $\mathcal{C}_u$ is constructed from the neighboring APs. From the AP side, the users to be served by AP $b$ are represented through the set $\mathcal{E}_b$. The sets $\{\mathcal{E}_b : b \in\mathcal{B}\}$ can be directly obtained from the sets $\{\mathcal{C}_u : u \in \mathcal{U}\}$, where $u \in \mathcal{E}_b \Leftrightarrow b \in \mathcal{C}_u$. Due to user-centric clustering, we use global indices $b$ and $u$ to refer to the APs and the typical user, respectively.
	
\begin{figure}[t]
		\centering
		\includegraphics[width=1.05\linewidth]{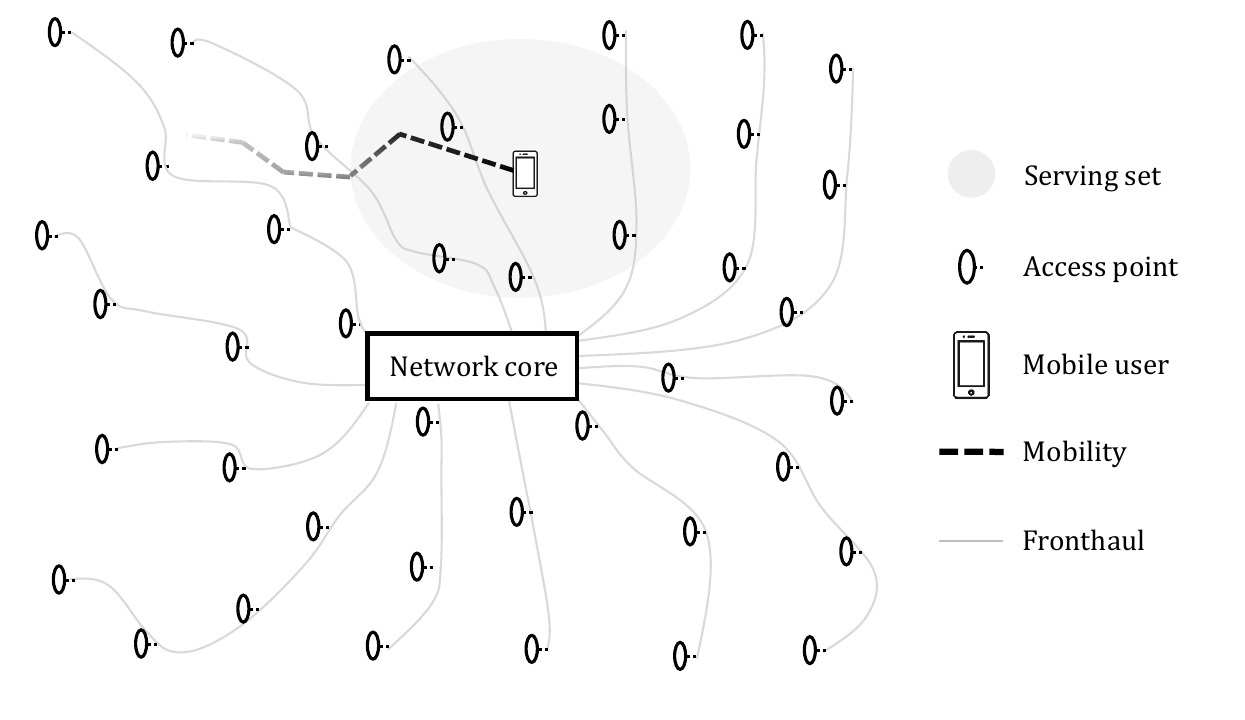}
		\vspace{-2em}
		\caption{User-centric cell-free MIMO network depicting a moving user.}
		\label{fig:UserCentricClustering}
		\vspace{-1em}
	\end{figure}

	Fig.~\ref{fig:UserCentricClustering} depicts our network model, where we show a mobile user. 
	We assume that the APs are connected to the network core through an unlimited capacity wired fronthaul network. A high bandwidth fiber optics or the radio stripes system~\cite{frenger2022antenna} are candidates to build this fronthaul network. This assumption is critical in our study to focus on the wireless aspects of the HO problem, and hence ignore the effect of the HO delays and the overheads of control signals on the performance. 
	Nonetheless, it is worth noting that our POMDP-based HO approach derived next can work with partial observable LSF statistics, which is an advantage. Moreover, as mentioned in the contributions, our approach takes into account future rewards to derive a HO policy with a manageable complexity. As we will see, our proposed approach is based on a rate metric that can be modified to take into account the limitations imposed by the fronthaul.
	
	In the following subsections, we develop the main elements needed for a proper formulation for the HO problem as a POMDP. Basically, we define a formula for the data rate that takes into account the undesired effect of mobility, i.e., channel aging, and the limited CSI availability. This formula will be used to define the reward function in the POMDP framework.
	
	\subsection{Channel Aging Model}
	User-mobility causes temporal variations in the channels between the user and the APs leading to a time-varying channel model even within the conventional channel coherence interval denoted as $\tau_{\rm c}$. At any time index $n = 0, 1, 2, \dots, \tau_{\rm c}-1$ inside a time window of duration $\tau_{\rm c}$, the channel realization between AP $b$ and user $u$ is modeled as ${\bf h}_{bu}[n] \triangleq \sqrt{ \beta_{bu} } {\bf g}_{bu}[n] \in \mathbb{C}^{M \times 1}$, where ${\bf g}_{bu}[n] \sim \mathcal{CN}\left({\bf 0}, {\bf I}_M \right)$ is the small-scale fading, and $\beta_{bu}$ is the LSF that accounts for the shadowing and the path loss. 
	The elements of the vector ${\bf g}_{bu}[n]$ are wide-sense stationary (WSS) processes because both the first moment and the autocovariance do not change with respect to the considered time window; also the second moment is always finite~\cite{CSIAging6608213}. Moreover, $\beta_{bu}$ is constant inside the considered time window because it changes very slowly compared to the duration $\tau_{\rm c}$ even for a high-speed mobile user. However this is not the case for ${\bf g}_{bu}[n]$, and hence the usage of the index notation ${\bf g}_{bu}[n]$. For any two time instants $n$ and $n'$, ${\bf g}_{bu}[n]$ and ${\bf g}_{bu}[n']$ are correlated~\cite{CSIAgingzheng2020cell}, i.e., $\{{\bf g}_{bu}[n] : \forall n\}$ are temporally correlated. 
	
	Let ${\bf C}_{bm,u}$ be the temporal covariance matrix of the channel components between antenna $m$ of AP $b$ and user $u$ at the considered instants, i.e, across $[g_{bm,u}[0], g_{bm,u}[1], \dots, g_{bm,u}[\tau_{\rm c}-1]]$. We model the change (aging) between time instants $n$ and $n'$ according to Jakes' model~\cite{jakes1994microwave}
	\begin{align}\label{eq:aging}
		[{\bf C}_{bm,u}]_{n n'} &= \mathbb{E}\left\{ g_{bm,u}[n] g_{bm,u}[n'] \right\} = \rho_{u}[n - n']
		\nonumber
		\\
		& \triangleq J_0 \left( 2 \pi \left( n - n' \right) f_{{\rm{D}}_u} T_{\rm s} \right)
	\end{align}
	where 
	$J_0(\cdot)$ is the zeroth-order Bessel function of the first kind, $T_{\rm s}$ is the sampling period of each channel use, and $f_{{\rm{D}}_u} = \frac{\mathtt{v}_u}{\rm \lambda_0}$ is the maximum Doppler shift which depends on the mobility relative speed $\mathtt{v}_u$ of the user and the carrier wavelength ${\rm \lambda_0}$. 
	
	In our channel model, within each duration $\tau_c$, we are interested in relating the channel realization at any time instant $n$ to the channel at the reference time instant 
	at which we have performed channel estimation. Based on this intuition and the Jakes model in~\eqref{eq:aging}, for each communication cycle of length $\tau_{\rm c}$, the small-scale fading at instant $n$ can be written as a function of an initial state ${\bf g}_{bu}[0]$ and an innovation component ${\bf v}_{bu}$ as follows~\cite{channelAgingMassiveMIMO8122014}
	\begin{align}\label{eq:channelEvolution}
		{\bf g}_{bu}[n] = \rho_{u}[n] {\bf g}_{bu}[0] + \bar{\rho}_{u}[n] {\bf v}_{bu}[n],
	\end{align}
	where ${\bf v}_{bu}[n] \sim \mathcal{CN}\left({\bf 0}, {\bf I}_M \right)$ is the independent innovation component at instant $n$, $\rho_{u}[n]$, given in~\eqref{eq:aging}, is the temporal correlation coefficient of user $u$ between channel realizations at instants $0$ and $n$, with $0 \le \rho_{u}[n] \le 1$, and $\bar{\rho}_{u}[n] = \sqrt{ 1 - |\rho_{u}[n]|^2 }$.

	The correlated small-scale fading model in~\eqref{eq:channelEvolution} is only needed within each conventional channel coherence duration $\tau_{\rm c}$ to relate the channel at the moment of estimation to other moments. Whether the small-scale fading between each coherence interval are dependent or not will not affect the analysis. However, for the large-scale fading, this is not true and a correlated shadowing model is needed for an accurate analysis. In this regard, as the user moves, there are common structures (e.g., buildings, walls) that will lead to a correlated shadowing. The large-scale fading model will be introduced later in the paper.
	\vspace{-1em}
	\subsection{Channel Estimation}\label{subsection:ChanEstimation}
	Channel aging also affects the channel instants at which the pilot sequence was transmitted. We believe this is more realistic than the assumption in~\cite{ChAgingPhaseNoise9471851}, where the channel is assumed to remain constant during the training phase. We note that~\cite{ChAgingPhaseNoise9471851} does not study HO. We assume that the set $\mathcal{U}_{i}$ represents users transmitting their pilot in time index $i$, such that $1 \le i \le \tau_{\rm p}$, where $\tau_{\rm p}$ is the length of the pilot training phase. 
	We define the pilot at channel use $n \le \tau_{\rm p}$ for each user $u \in \mathcal{U}_{i}$ as $\phi_u[n] = \delta[n - i]$, with $\delta[n]$ the Kronecker delta function.
	
	At instant $i$ during the uplink pilot training phase, i.e., $i \le \tau_{\rm p}$, the signal received at AP $b$ is
	\begin{align}\label{eq:Signal_at_instant_i}
		{\bf y}_{b}[i]
		&= \sum_{u' \in \mathcal{U}_i} \sqrt{p^{({\rm u})}} 
		{\bf h}_{bu'}[i]
		+ 
		{\bf z}_{b}[i]
	\end{align}
	where $p^{({\rm u})}$ is the uplink transmit power of the pilot signal, and ${\bf z}_{b}$ is the noise with entries distributed as $\mathcal{CN}(0, \sigma_{\rm z}^2)$.
	
	Based on~\eqref{eq:channelEvolution}, ${\bf g}_{bu}[n]$ at time instant $n \le \tau_{\rm p}$ can be related to ${\bf g}_{bu}[n_{\rm est}]$ at time instant $n_{\rm est} = \tau_{\rm p} + 1$, i.e., at the time of estimating the channel, as follows:
	\vspace{-0.5em}
	\begin{align}\label{eq:channelEvolutionChanEst}
		{\bf g}_{bu}[n] = \rho_{u}[n_{\rm est} - n] {\bf g}_{bu}[n_{\rm est}] + \bar{\rho}_{u}[n_{\rm est} - n] {\bf v}_{bu}[n]
	\end{align}
	
	Using~\eqref{eq:channelEvolutionChanEst} in the small-scale fading component of the channel ${\bf h}_{bu}[i]$ in~\eqref{eq:Signal_at_instant_i}, and assuming that AP $b$ estimates the channel for user $u \in \mathcal{U}_i$, the signal received at AP $b$ at instant $i$ during the uplink pilot training phase can be represented as
	\begin{align}\label{eq:Signal_at_instant_i_detailed}
		{\bf y}_{bu}[i]
		&=
		\resizebox{0.2\textwidth}{!}
		{$\displaystyle
			\sqrt{p^{({\rm u})}} 
			\rho_{u}[n_{\rm est} - i]
			{\bf h}_{bu}[n_{\rm est}] 
			$} 
			+
			\resizebox{0.19\textwidth}{!}
			{$\displaystyle
			\sqrt{p^{({\rm u})} \beta_{bu}}
			\bar{\rho}_{u}[n_{\rm est} - i] {\bf v}_{bu}[i]
			$}
			\nonumber \\
			&\quad
			+
			\resizebox{0.175\textwidth}{!}
			{$\displaystyle
				\sum_{u' \in \mathcal{U}_i, u' \ne u} \sqrt{p^{({\rm u})}} 
			{\bf h}_{bu'}[i]
			$}
			+ 
			{\bf z}_{b}[i]
	\end{align}
	
	Using linear minimum mean square error (LMMSE), AP $b$ estimates the channel of user $u \in \mathcal{U}_i$~as
	\begin{align}\label{eq:est_chan}
		{\bf \hat{h}}_{bu}[n_{\rm est}]
		=
		\frac{ \rho_{u}[n_{\rm est} - i ] \sqrt{p^{({\rm u})}} \beta_{bu}}{ \sum_{u' \in \mathcal{U}_i} p^{({\rm u})} \beta_{bu'} + \sigma_{\rm z}^2 }
		{\bf y}_{bu}[i],
		\quad \text{for}\ u \in \mathcal{U}_i
	\end{align}
	When using LMMSE estimation, the channel estimation error $\mathrm{ {\bf e}}_{bu}[n_{\rm est}] = {\bf h}_{bu}[n_{\rm est}] - {\bf \hat{h}}_{bu}[n_{\rm est}]$ is uncorrelated with the estimated channel ${\bf \hat{h}}_{bu}[n_{\rm est}]$ and is distributed as $\mathrm{ {\bf e}}_{bu}[n_{\rm est}] \sim \mathcal{CN}\left({\bf 0}, {\bm \Theta_{bu}}\right)$, where the covariance ${\bm \Theta_{bu}} \triangleq \beta_{bu} {\bf I}_M - \psi_{bu} {\bf I}_M$, with
	\begin{align}\label{eq:var_estCh}
		\psi_{bu}[n_{\rm est}]
		=
		\frac{\rho_{u}^2[ n_{\rm est} - i  ] p^{({\rm u})} \beta_{bu}^2 }{ \sum_{u' \in \mathcal{U}_i} p^{({\rm u})} \beta_{bu'} + \sigma_{\rm z}^2 },
		\quad \text{for}\ u \in \mathcal{U}_i
	\end{align}

	\vspace{-1em}
	\subsection{Downlink Signal Model and Lower-Bound for Achievable Rate}
	The APs use conjugate beamforming to serve users. Although conjugate beamforming is not the optimal beamformer choice~\cite{ammar9650567}, it is used in cell-free MIMO schemes to allow for ease of implementation. In addition, conjugate beamforming does not impose an overhead on the control plane of the fronthaul in terms of requiring the exchange of CSI.
	
	Using conjugate beamforming, each AP uses the channels estimated at time instant $n_{\rm est}$ to construct the beamformer and transmit data to its users $\mathcal{E}_b$ within the time instants $n_{\rm est} \le n \le \tau_c$. The signal sent by AP $b$ at time instant $n \ge n_{\rm est}$ to users $\mathcal{E}_b$ can be written as
	\begin{align}
		{\bf x}_b[n] =
		\sum_{u \in \mathcal{E}_b}
		\sqrt{\eta_{bu}[n_{\rm est}]}
		{\bf \hat{h}}_{bu}^*[n_{\rm est}] s_{u}[n],
	\end{align}
	where ${\bf \hat{h}}^*[n_{\rm est}]$ is the conjugate of the channel estimated at time instant $n_{\rm est}$, $s_{u}[n]$ is the complex data symbol for the user with $\mathbb{E}\{|s_{u}[n]|^2\} = 1$, and $\eta_{bu}[n_{\rm est}]$ is the transmit power allocated by AP $b$ for user $u$ to satisfy the statistical power budget~\cite{9141340} defined as $\mathbb{E}\{\| {\bf x}_b[n] \|^2\} \le p^{({\rm d})}$, 
	and it is defined~as
	\begin{align}\label{eq:eta}
		\eta_{bu}[n_{\rm est}] =
		\frac{p^{({\rm d})}}
		{
			M |\mathcal{E}_b| \mathbb{E}\{\|\hat{\bf h}_{bu}[n_{\rm est}]\|^2\}
		}
		=
		\frac{p^{({\rm d})}}
		{
			M |\mathcal{E}_b| \psi_{bu}[n_{\rm est}]
		},
	\end{align}
	where $p^{({\rm d})}$ is the power budget of AP~$b$.
	
	As discussed previously, channel aging affects even the data transmission phase. Using~\eqref{eq:channelEvolution} to model channel evolution inside a communication cycle, the channel at time instant $n \ge n_{\rm est}$ can be represented as ${\bf h}_{bu}[n] = \sqrt{\beta_{bu}} {\bf g}_{bu}[n]$, with
	\begin{align}\label{eq:channelEvolution_data}
		{\bf g}_{bu}[n] = \rho_{u}[n - n_{\rm est}] {\bf g}_{bu}[n_{\rm est}] + \bar{\rho}_{u}[n - n_{\rm est}] {\bf v}_{bu}[n]
	\end{align}
	
	The signal received at user $u$ during the data transmission phase, i.e., at $n \ge n_{\rm est}$, is shown in~\eqref{eq:signalModel}, 
\begin{figure*}[b]
	\hrule
		\begin{align}\label{eq:signalModel}
			&y_{u}[n] =
			\sum_{b \in \mathcal{B}} 
			{\bf h}_{bu}^T[n] {\bf x}_b[n]
			+ z_u
			\nonumber \\[-5pt]
			&\quad =
			\sum_{b \in \mathcal{C}_u} \sqrt{\eta_{bu}[n_{\rm est}] \beta_{bu}} 
			\left(
			\rho_{u}[n - n_{\rm est}] {\bf g}_{bu}^T[n_{\rm est}] + \bar{\rho}_{u}[n - n_{\rm est}] {\bf v}_{bu}^T[n]
			\right)	
			{\bf \hat{h}}_{bu}^*[n_{\rm est}]
			s_{u}[n]
			\nonumber \\[-5pt]
			& \quad \quad
			+
			\sum_{u' \in \mathcal{U}, u' \ne u}
			a_{uu'}[n] s_{u'}[n]
			+ z_u
			\nonumber \\[-5pt]
			& \quad =
			\underbrace{
				\rho_{u}[n - n_{\rm est}]
				\sum_{b \in \mathcal{C}_u} \sqrt{\eta_{bu}[n_{\rm est}]} 
				\mathbb{E}\left\{
				{\bf h}_{bu}^T[n_{\rm est}]
				{\bf \hat{h}}_{bu}^*[n_{\rm est}]
				\right\}
				s_{u}[n]
			}_{\mathrm{desired~signal}~{\rm DS}_u[n]}
			\nonumber \\[-5pt]
			& \quad \quad
			+
			\underbrace{
				\rho_{u}[n - n_{\rm est}]
				\sum_{b \in \mathcal{C}_u} \sqrt{\eta_{bu}[n_{\rm est}]}  
				\left(
				{\bf h}_{bu}^T[n_{\rm est}]
				{\bf \hat{h}}_{bu}^*[n_{\rm est}]
				-
				\mathbb{E}\left\{
				{\bf h}_{bu}^T[n_{\rm est}]
				{\bf \hat{h}}_{bu}^*[n_{\rm est}]
				\right\}
				\right)
				s_{u}[n]
			}_{\mathrm{beamformer~uncertainity}~{\rm BU}_u[n]}
			\nonumber \\[-5pt]
			& \quad \quad
			+
			\underbrace{
				\bar{\rho}_{u}[n - n_{\rm est}]
				\sum_{b \in \mathcal{C}_u} \sqrt{\eta_{bu}[n_{\rm est}] \beta_{bu}} 
				{\bf v}_{bu}^T[n]
				{\bf \hat{h}}_{bu}^*[n_{\rm est}]
				s_{u}[n]
			}_{\mathrm{channel~aging}~{\rm CA}_u[n]}	
			+
			\sum_{u' \in \mathcal{U}, u' \ne u}
			\!\!\!\!
			\underbrace{
				a_{uu'}[n] s_{u'}[n]
			}_{\mathrm{multiuser~interference~{\rm MI}_u[n]}}
			\!\!\!\!\!\!
			+ \underbrace{z_{u}}_{\mathrm{noise}},
	\end{align}
\vspace{-2em}
\end{figure*}
where $a_{uu'}[n] = \sum_{b' \in \mathcal{C}_{u'}} \sqrt{\eta_{b'u'}[n_{\rm est}]} {\bf h}_{b'u}^T[n] \hat{\bf h}_{b'u'}^{*}[n_{\rm est}]$. Using the signal model in~\eqref{eq:signalModel}, 
we can characterize the performance through a lower bound for the channel capacity, which is defined as:
\begin{align}\label{eq:rate_LB}
	R_u^{(\rm lb)} &=
	\frac{1}{\tau_{\rm c}} \sum_{n=n_{\rm est}}^{\tau_{\rm c}} \log \left( 1 + \frac{ \left| \mathbb{E}\left\{{\rm DS}_u[n]\right\} \right|^2 }{ A_u [n] } \right)
	\end{align}
	with
	\begin{align}
		A_u [n] &=
		\mathbb{E}\left\{\left|{\rm BU}_u[n]\right|^2\right\}
		+ \mathbb{E}\left\{\left| {\rm CA}_u[n] \right|^2\right\}
		\nonumber \\
		&\quad
		+ \sum_{u' \in \mathcal{U}, u' \ne u}
		\mathbb{E}\left\{\left| {\rm MI}_{uu'}[n] \right|^2\right\}
		+ \sigma_{\rm z}^2
		\nonumber \\
		&=
		\xi_{2,3,u}[n] + \sum_{u' \in \mathcal{U}, u' \ne u} \xi_{4,uu'}[n] + \sigma_{\rm z}^2
	\end{align}
	where $\log$ is the natural logarithm, and 
	the powers of desired signal, beamformer uncertainty + channel aging, and interference can be, respectively, written in closed-form~as
	{\allowdisplaybreaks
		\begin{align}
			\xi_{1,u}[n] & =
			M p^{({\rm d})}
			\rho_{u}^2[n - n_{\rm est}]
			\left|
			\sum_{b \in \mathcal{C}_u} \sqrt{
				\frac{\psi_{bu}[n_{\rm est}]}{|\mathcal{E}_b|}
			} 
			\right|^2
			%
			%
			\\[-5pt]
			\xi_{2,3,u}[n] & =
			M
			\sum_{b \in \mathcal{C}_u} 
			\frac{p^{({\rm d})} \beta_{bu} }
			{
				|\mathcal{E}_b|
			}
			%
			%
			\\[-5pt]
			\xi_{4,uu'}[n] & = 
			\nonumber \\
			&
			\hspace{-3em}
			\begin{cases}
				\begin{aligned}
					&
					M \bigg(\sum_{b' \in \mathcal{C}_{u'}} \frac{p^{({\rm d})}}
					{
						|\mathcal{E}_{b'}|
					}
					\beta_{b'u}
					\\
					&\quad \quad 
					+
					\rho_{u}^2[n - n_{\rm est}]
					\bigg|\sum_{b' \in \mathcal{C}_{u'}} \sqrt{\frac{p^{({\rm d})} \psi_{b'u} }
						{
							|\mathcal{E}_{b'}|
						}
					}
					\bigg|^2
					\bigg)
				\end{aligned}
				,& \text{if}\ u' \in \mathcal{U}_i
				\\
				\displaystyle
				M \sum_{b' \in \mathcal{C}_{u'}} 
				\frac{p^{({\rm d})}}
				{
					|\mathcal{E}_{b'}|
				}
				\beta_{b'u}
				,& \text{if}\ u' \notin \mathcal{U}_i
			\end{cases}
	\end{align}}
	\vspace{-1em}
	\begin{proof}
		The proof for~\eqref{eq:rate_LB} follows similar steps as in~\cite{CSIAgingzheng2020cell} with some minor differences. To make the discussion self-contained, the details of the derivations are provided in Appendix~\ref{Appendix:rate_LB}.
	\end{proof}

	\vspace{-1em}
	\section{Problem Formulation using POMDP}\label{sec:POMDP}
	%
	
	To study HOs, we develop a solution that focuses on each user independently
	; we denote this user as the typical user $u$. The movement necessitates that each user considers a HO that is described through changing the serving set $\mathcal{C}_u$ between communication cycles $(t-1)$ and $t$.
	
	\vspace{-1em}
	\subsection{Components of the POMDP}\label{subsec:POMDP_compon}
	The channels from one communication cycle (of length $\tau_{\rm c}$) to another evolves over time as a finite-state Markov process. 
	We will use the notation $(\cdot)^{(t)}$ to refer to the variables in communication cycle $t$ (equivalently, decision cycle). Each cycle is of length $\tau_c$ channel uses. We also make use of notation to define $\mathcal{E}_b^{(t)}$ and $\mathcal{C}_u^{(t)}$ as the users served by AP $b$ and as the APs serving user $u$, respectively, at decision cycle $t$. These variables will change if a HO is~initiated.
	
	
	We define our POMDP framework using the following tuple
	\vspace{-0.5em}
	\begin{align}\label{eq:POMDP_model}
		&\mathcal{P}(\mathcal{B}, T_{\rm H})
		=
		\nonumber \\
		&
		\big(\mathcal{S}(|\mathcal{B}|), \mathcal{A}(|\mathcal{B}|), \Omega(|\mathcal{B}|), {\bf P}_{\rm s, \mathcal{B}}^{(t)}
		, {\bf P}_{\rm o, \mathcal{B}}^{(t)},
		{\bm \omega}^{(0)}_{\mathcal{B}}, R_{\mathcal{B}}\left({\bf s}^{(t)}, {\bf a}^{(t)}\right), T_{\rm H} \big)
		,
	\end{align}
	where $\mathcal{S}(|\mathcal{B}|)$ denotes the state space corresponding to the AP set $\mathcal{B}$, $\mathcal{A}(|\mathcal{B}|)$ is the action space, $\Omega(|\mathcal{B}|)$ is the observation space, ${\bf P}_{\rm s, {\mathcal{B}}}^{(t)}$ is the transition probability matrix of the states, ${\bf P}_{\rm o, \mathcal{B}}^{(t)}$ is the observation probability matrix
	, ${\bm \omega}^{(0)}_{\mathcal{B}}$ is the initial belief distribution, $R_{\mathcal{B}}\left({\bf s}^{(t)}, {\bf a}^{(t)}\right)$ is the reward function which depends on the action ${\bf a}^{(t)}$ and state ${\bf s}^{(t)}$ found at decision cycle $t$, and $T_{\rm H}$ is the time horizon which allows us to consider future rewards when taking our actions. 
	We will use the indices $i$, $j$ and $l$ to refer to a specific state vector $\widetilde{\bf s}_i$, action vector $\widetilde{\bf a}_j$ and observation vector $\widetilde{\bf o}_l$, respectively. Notations $\mathcal{S}(|\mathcal{B}|), \mathcal{A}(|\mathcal{B}|), \Omega(|\mathcal{B}|)$ and $\mathcal{S}, \mathcal{A}, \Omega$ are used interchangeably.
	
	
	
	\subsubsection{State space $\mathcal{S}(|\mathcal{B}|)$}
	The state space represents all the possible combinations of the states of the channels between the APs and the users. Specifically, we represent the state of the channel between the user and AP $b$ at decision cycle $t$ through the scalar $s_{bu}^{(t)} \in \mathcal{B}$ defined as a discrete version of the LSF. The value of the LSF is quantized into $Q$ levels. Then, the combination of the channel states for all the APs comprise the POMDP state (or simply the state) represented by the vector ${\bf s}^{(t)} = [\{s_{bu}^{(t)} : b \in \mathcal{B} \}]^T \in \mathbb{R}^{|\mathcal{B}| \times 1}$.
		
	To prevent confusion, ${\bf s}^{(t)}$ simply represents the POMDP \emph{state} (or simply state) at decision cycle $t$, while $s_{bu}^{(t)}$ represents a single \emph{channel state}. If we have $Q$ possible values for $s_{bu}^{(t)}$, then we have $Q^{|\mathcal{B}|}$ 
	possible states.

	%
	%
	%
	
	\subsubsection{Action space $\mathcal{A}(|\mathcal{B}|)$}
	It represents all the possible combinations of the different actions. At a specific decision cycle $t$, a user $u$ can either connect to AP $b \in \mathcal{B}$ or not. We define $a_{bu}^{(t)} \in \{0, 1\}$ to represent the decision of either connect ($a_{bu}^{(t)} = 1$) or not connect ($a_{bu}^{(t)} = 0$) at $t$. We use the notation ${\bf a}^{(t)} = [\{a_{bu}^{(t)} : b \in \mathcal{B} \}]^T \in \mathbb{B}^{|\mathcal{B}| \times 1}$ to denote the action for user $u$ at $t$ with the APs, i.e., the vector ${\bf a}^{(t)}$ represents the action decided at $t$ by the POMDP policy.
	
	We further assume that a user will be connected to $B_{\rm con} < |\mathcal{B}|$ APs. This is represented through the condition $\sum_{b \in \mathcal{B}} a_{bu}^{(t)} = B_{\rm con}$. Using a fixed $B_{\rm con}$ allows us to shrink the action space compared to an unconstrained number of associations. Based on this, $\mathcal{A}(|\mathcal{B}|)$ contains ${|\mathcal{B}| \choose B_{\rm con} } = \frac{|\mathcal{B}|!}{B_{\rm con}! \left(|\mathcal{B}| - B_{\rm con} \right)!}$ 
	possible actions. The action space can be very huge if we consider that the user can connect to an unconstrained number of APs, hence, placing this constraint is crucial to manage the complexity of the model.
	
	
	
	\subsubsection{Observation Space $\Omega(|\mathcal{B}|)$} The vector ${\bf o}^{(t)} = [\{o_{bu}^{(t)} : b \in \mathcal{B} \}]^T \in \mathbb{R}^{|\mathcal{B}| \times 1}$ is used to represent the observation at decision cycle $t$. We assume that the channel state between the user and the \emph{currently connected} APs is known, i.e., if $a_{bu}^{(t)} = 1$, then $o_{bu}^{(t)} = s_{bu}^{(t)}$ and the channel state is observable. If the user is not connected to AP $b$, then the channel state $s_{bu}^{(t)}$ is \emph{not observable} and this will affect the observability of the POMDP state ${\bf s}^{(t)}$. In such a case, a probabilistic approach should be constructed, which we will discuss later in this paper. 
	
	\subsubsection{Transition probability} 
	We use ${\bf P}_{\rm s, {\mathcal{B}}}^{(t)} \in \mathbb{R}^{|\mathcal{S}| \times |\mathcal{S}|}$ to denote the transition probability matrix for the POMDP states at decision cycle $t$. Each element $[{\bf P}_{\rm s, {\mathcal{B}}}^{(t)}]_{i,i'} \in [0,1]$ (at row $i$ and column $i'$ of matrix ${\bf P}_{\rm s, {\mathcal{B}}}^{(t)}$) is defined as
	\begin{align}\label{eq:Transition_prob}
		%
		[{\bf P}_{\rm s, {\mathcal{B}}}^{(t)}]_{i,i'}
		=
		\mathbb{P}({\bf s}^{(t)} = \widetilde{\bf s}_{i'}|\ {\bf s}^{(t-1)} = \widetilde{\bf s}_i), \quad \tilde{\bf s}_i, \tilde{\bf s}_{i'} \in \mathcal{S}(|\mathcal{B}|)
	\end{align}
	As can be seen from~\eqref{eq:Transition_prob}, the evolution of states is independent from the action ${\bf a}^{(t-1)}$ taken at decision cycle $t$. Each row in the matrix ${\bf P}_{\rm s, {\mathcal{B}}}^{(t)}$ sums up to $1$.
	

	\subsubsection{Observation distribution} 
	We use ${\bf P}_{\rm o, \mathcal{B}}^{(t)} \in \mathbb{R}^{|\mathcal{S}|  \times |\Omega| \times |\mathcal{A}|}$ to represent the matrix of the observation probabilities. 
	Herein, we assume that we can only observe the CSI for the APs that the user is currently connected to. Element $\left[{\bf P}_{\rm o, \mathcal{B}}^{(t)}(\widetilde{\bf o}_l)\right]_{i, l,j} \in [0,1]$ is defined as
	\begin{align}
		\left[{\bf P}_{\rm o, \mathcal{B}}^{(t)}\right]_{i,l,j}
		&=
		\mathbb{P}({\bf o}^{(t)} = \widetilde{\bf o}_l|\ {\bf s}^{(t)} = \widetilde{\bf s}_i, {\bf a}^{(t)} = \widetilde{\bf a}_j);
		\nonumber \\
		& \quad
		\widetilde{\bf s}_i \in \mathcal{S}(|\mathcal{B}|),\  \widetilde{\bf o}_l \in \Omega(|\mathcal{B}|),\ \widetilde{\bf a}_j \in \mathcal{A}(|\mathcal{B}|)
	\end{align}
	
	%
	%
	%
	%
	%
	%
	
	\subsubsection{Belief} The belief denotes the posterior probability distribution over the state space, and it represents the knowledge of the decision maker about the state of the POMDP based on the past actions and observations. 
	We use ${\bm \omega}^{(t)}_{\mathcal{B}} = [\omega_{1, \mathcal{B}}^{(t)}, \dots, \omega_{|\mathcal{S}|, \mathcal{B}}^{(t)}]$ as the belief vector at decision cycle $t$, where $\omega_{i, \mathcal{B}}^{(t)}$ is the probability of ${\bf s}^{(t)}$ to equal a particular value $\widetilde{\bf s}_i \in \mathcal{S}$, given all the action and observation history from $t=0$ till decision cycle $(t-1)$. We denote this history as $\mathcal{H}_{t-1} = \{ {\bf o}^{(t-1)}, {\bf a}^{(t-1)}, \mathcal{H}_{t-2}\}$. 
	Mathematically, we define $\omega_{i, \mathcal{B}}^{(t)}$~as
	\begin{align}
		\omega_{i, \mathcal{B}}^{(t)} = \mathbb{P}\left( {\bf s}^{(t)} = \widetilde{\bf s}_i |\ \mathcal{H}_{t-1}\right);
		\quad \widetilde{\bf s}_i \in \mathcal{S}(|\mathcal{B}|)
	\end{align}
	It is worth noting that the belief state $\omega_{i, \mathcal{B}}^{(t)}$ is a sufficient statistic for the history $\mathcal{H}_{t-1}$~\cite{smallWoodSondik}, because once the belief is known, we do not need $\mathcal{H}_{t-1}$ to take decisions.
	
	\subsubsection{Reward} $R_{\mathcal{B}}({\bf s}^{(t)}, {\bf a}^{(t)})$ denotes the reward when executing action ${\bf a}^{(t)}$ at POMDP state ${\bf s}^{(t)}$. We define the POMDP formulation independently for each user using the typical user notation, hence, we use a single-user version of the spectral efficiency derived in~\eqref{eq:rate_LB} to define the reward for the POMDP, i.e., an SNR-based spectral efficiency. 
	Specifically, we make use of the notation above to rewrite the variance of the estimated channel in~\eqref{eq:var_estCh} for a single-user case as
	\begin{align}\label{eq:psi_POMDP_SingleU}
		\psi_{{\rm S}, bu}[s_{bu}^{(t)}]
		=
		\frac{\rho_{u}^2[ n_{\rm est} - i ] p^{({\rm u})} \big(s_{bu}^{(t)}\big)^2 }{ \sigma_{\rm z}^2 },
		\quad \text{for}\ u \in \mathcal{U}_i
		.
	\end{align}
	In turn, the allocated power $\eta_{{\rm S}, bu}[n_{\rm est}]$ to user $u$ becomes
	\begin{align}
		\eta_{{\rm S}, bu}[n_{\rm est}] =
		\frac{p^{({\rm d})}}
		{
			M |\mathcal{E}_b| \psi_{{\rm S}, bu}[s_{bu}^{(t)}]
		}.
	\end{align}

	Moreover, we use the fact that we can relate any term $c_{bu}$ that contains a summation over $\mathcal{C}_u^{(t)}$ to a summation over all the APs $\mathcal{B}$ through
	\begin{align}\label{eq:action_POMDP}
		\sum_{b \in \mathcal{C}_u^{(t)}} c_{bu}
		=
		\sum_{b \in \mathcal{B}} a_{bu}^{(t)} c_{bu}
	\end{align}
	
	Using~\eqref{eq:psi_POMDP_SingleU} and~\eqref{eq:action_POMDP} in~\eqref{eq:rate_LB}, we define the reward used for the POMDP formulation as: 
	\begin{align}\label{eq:reward_SingleU}
		R_{\mathcal{B}}\left({\bf s}^{(t)}, {\bf a}^{(t)}\right) =
		\frac{1}{\tau_{\rm c}} \sum_{n=n_{\rm est}}^{\tau_{\rm c}} \log \left( 1 + \frac{ 
		\resizebox{0.28\linewidth}{!}
		{$
		\xi_{1,{\rm S}, u}\left[n, {\bf s}^{(t)}, {\bf a}^{(t)}\right]
		$}
		}{\displaystyle
		\resizebox{0.36\linewidth}{!}
		{$
			\xi_{2,3,{\rm S}, u}\left[n, {\bf s}^{(t)}, {\bf a}^{(t)}\right] + \sigma_{\rm z}^2
		$}	
		}	
		\right) 
	\end{align}
	where
	\vspace{-1em}
	\begin{align}
		\xi_{1,{\rm S}, u}\left[n, {\bf s}^{(t)}, {\bf a}^{(t)}\right] & =
		\resizebox{0.61\linewidth}{!}
		{$\displaystyle
		M p^{({\rm d})}
		\rho_{u}^2[n - n_{\rm est}]
		\left|
		\sum_{b \in \mathcal{B}}
		a_{bu}^{(t)}
		\sqrt{
			\frac{\psi_{{\rm S}, bu} [s_{bu}^{(t)} ]}{|\mathcal{E}_b|}
		} 
		\right|^2
		$}
		\\
		\xi_{2,3,{\rm S}, u}\left[n, {\bf s}^{(t)}, {\bf a}^{(t)}\right] & =
		M
		p^{({\rm d})}
		\sum_{b \in \mathcal{B}} 
		\frac{a_{bu}^{(t)} s_{bu}^{(t)} }
		{
			|\mathcal{E}_b|
		}
	\end{align}

	Using long-term statistics for taking HO decisions is more reliable than instantaneous channel realizations which fluctuate rapidly and could lead to frequent unnecessary HOs that disrupt the user's experience. Long-term statistics allow for strategic HOs decisions, and it helps filter out momentary disruptions in the channels.


	\subsubsection{Time horizon $T_{\rm H}$} The time horizon is the number of decision cycles in the future to consider when optimizing the cumulative expected discounted
	reward of the POMDP.

	\vspace{-1em}
	\subsection{Objective}
\begin{figure}[t]
		\centering
		\includegraphics[width=1\linewidth]{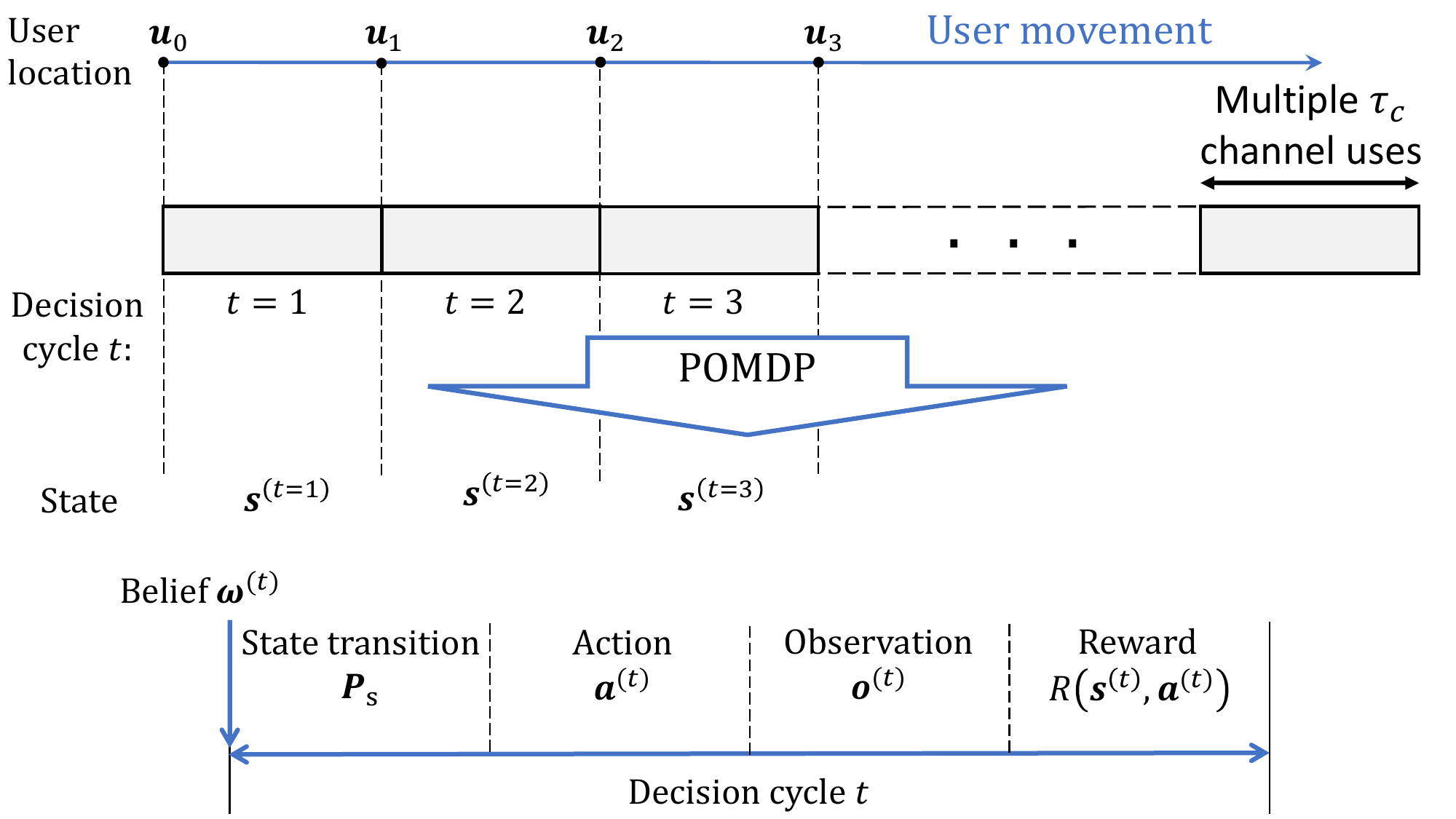}
		\vspace{-1.4em}
		\caption{HO procedure as a POMDP model.}
		\label{fig:mob_POMDPform}
		\vspace{-1.5em}
	\end{figure}

	\vspace{-0.5em}
	A key aspect for the POMDP model is the partial observability of the states. Hence, the states cannot be directly observed, and instead we have access to observation ${\bf o}^{(t)}$ that gives incomplete information about the current state ${\bf s}^{(t)}$. To deal with this problem, the belief vector ${\bm \omega}^{(t)}_{\mathcal{B}}$ is constructed to represent the internal belief for the state by the decision maker. 
	Hence, at each decision cycle $t$, the POMDP state ${\bf s}^{(t)}$ is updated. Using $\mathcal{H}_{t-1}$, the POMDP decision maker constructs a belief vector ${\bm \omega}^{(t)}_{\mathcal{B}}$ for the partially observable state and then performs the HO decisions for the user represented through the action ${\bf a}^{(t)}$. After executing action ${\bf a}^{(t)}$, we obtain the observation ${\bf o}^{(t)}$ and a reward $R_{\mathcal{B}}\left({\bf s}^{(t)}, {\bf a}^{(t)}\right)$. The same operation repeats for the next decision cycle. Fig.~\ref{fig:mob_POMDPform} illustrates this operation.

	We aim to construct a HO policy sequence $\Pi^\star$ that maps ${\bm \omega}^{(t)}_{\mathcal{B}}$ into action ${\bf a}^{(t)}$, i.e., ${\bf a}^{(t)} = \Pi^\star({\bm \omega}^{(t)}_{\mathcal{B}})$. For a finite horizon, the derived policy sequence is a set $\Pi^\star = \{\pi^{\star,(1)}, \pi^{\star,(2)}, \dots\}$ that contains policies specific for each decision cycle $t$, while for an infinite horizon, the obtained policy is stationary, i.e., we have a single policy $\Pi^\star = \pi^\star$, and the superscript $(t)$ can be dropped.
	
	
	The objective function of the POMDP is defined as the long term reward averaged over the different states as follows
	\begin{align}\label{eq:obj_POMDP}
		J_{\Pi}({\bm \omega}^{(0)}_{\mathcal{B}})
		=
		\mathbb{E}_{\mathcal{S}} \left\{ \sum_{t = 1}^{T_{\rm H}} \gamma^t R_{\mathcal{B}}\left({\bf s}^{(t)}, {\bf a}^{(t)}\right) | {\bm \omega}^{(0)}_{\mathcal{B}} \right\},
	\end{align}
	where $0\le \gamma < 1$ is a discount factor for future rewards, and it ensures that the sum is finite. The summation over $t$ in~\eqref{eq:obj_POMDP} goes to $T_{\rm H}$, which could be either a finite or infinite time horizon.
	
	
	The aim of the POMDP formulation is to determine an optimal policy $\Pi^\star$ that maximizes the objective function in~\eqref{eq:obj_POMDP}, i.e.,
	\begin{align}\label{eq:ObjPOMDP}
		\Pi^\star = \operatorname{arg}\ \underset{\Pi}{\operatorname{max}} \quad J_{\Pi}({\bm \omega}^{(0)}_{\mathcal{B}}),\ \text{for\ every}\ {\bm \omega}^{(0)}_{\mathcal{B}}
	\end{align}
	Once the formulation of the POMDP is complete, the value iteration algorithm~\cite{krishnamurthy2016partially, sondik1971optimal} can be used to obtain a policy sequence that determines the HO decisions of the user while moving.

	\vspace{-1.5em}
	\subsection{Location of Decision Maker}\label{sec:loc_decision_maker}
	\vspace{-0.3em}
	
	As noted earlier, a POMDP/MDP includes a feedback controller or a decision maker that receives observations from the Markov process and executes an action that is fed back to the process~\cite{krishnamurthy2016partially}. The decision maker will be the unit that will formulate the POMDP and derive the HO policy in~\eqref{eq:ObjPOMDP}.

	Based on our formulation, there are two candidate locations to deploy the decision maker. The first location is the central unit (CU) found at the network core, while the second location is the user. In the first option, the details needed to construct the POMDP should be forwarded from the APs to the CU. While in the second option, these details can be easily constructed at the user's side, which makes both deployments feasible.

	\vspace{-1em}
	
	\section{Problem Analysis and Simplification}\label{sec:Analysis}
	\vspace{-0.5em}
	So far, we formulated the main elements of the problem as a POMDP, however, the problem in its current form is too complex to be solved. The main issue resides in the large size of the transition and observation probability matrices and the computational complexity of solving the POMDP which can be $\mathcal{O}\left( |A| |\mathcal{S}|^2 \right)$ per iteration using the value iteration algorithm~\cite{condon1992complexity}. We note that $|\mathcal{S}| = Q^{|\mathcal{B}|}$, where, as defined earlier, $Q$ is the number of quantized levels for the LSF. The computational complexity is therefore a serious problem even for a reasonable number of APs.
	
	To overcome this issue, we propose a divide-and-conquer approach. Specifically, instead of constructing a single POMDP problem $\mathcal{P}(\mathcal{B}, T_{\rm H})$ with the set $\mathcal{B}$, we construct multiple POMDP sub-problems $\mathcal{P}(\mathcal{B}_{\rm cand, \ell}, T_{\rm H})$ with AP candidate sets $\mathcal{B}_{\rm cand, \ell} \subset \mathcal{B}; \forall \ell$. The set $\mathcal{B}_{\rm cand, \ell}$ is much smaller than $\mathcal{B}$ which allows us to define and solve the POMDP sub-problems without any issue of scalability. In our approach, each POMDP sub-problem is solved separately, and finally the POMDP sub-problem that produces the best total expected reward is selected, with its candidate APs and policy, to be used within a chosen time window. Details are provided in Section~\ref{sec:POMDP_subproblems}.
	
	\vspace{-1.2em}
	\subsection{Problem Analysis}
	\vspace{-0.5em}
	\subsubsection{State Levels}\label{sec:state_levels}
	We focus on the case of $Q = 2$, where as noted earlier, $Q$ is the number of quantization levels for the LSF $\beta_{bu}^{(t)}$. This implies that we classify the state $s_{bu}^{(t)}$ of any channel between AP $b$ and the typical user as either a good channel ($\widetilde{\beta}_{1}$) or a bad one ($\widetilde{\beta}_{0}$), i.e.,
	\begin{align}\label{eq:twoChannelState}
		s_{bu}^{(t)}
		=
		\begin{cases}
			\widetilde{\beta}_{1}, & \text{if}\ \beta_{bu}^{(t)} > \beta_{\rm threshold},
			\\
			\widetilde{\beta}_{0}, & \text{if}\ \beta_{bu}^{(t)} \le \beta_{\rm threshold}
		\end{cases}
	\end{align}
	As discussed in the definition of the state space in Section~\ref{sec:POMDP}, we use~\eqref{eq:twoChannelState} to construct the state as ${\bf s}^{(t)} = [\{s_{bu}^{(t)} : b \in \mathcal{B} \}]^T \in \mathbb{R}^{|\mathcal{B}| \times 1}$.

	The assumption of $Q=2$ helps to control the computational complexity to solve the POMDP; it provides a strong intuition where each channel is seen as either in a good or in a bad state. This assumption is further strengthened when we define the transition probability in a closed-form expression next, which characterizes how much the channel will stay or transition to any state in the future, and provides more distinction between the states of the different channels.
	
\begin{figure}[t]
		\centering
		\includegraphics[width=0.7\linewidth]{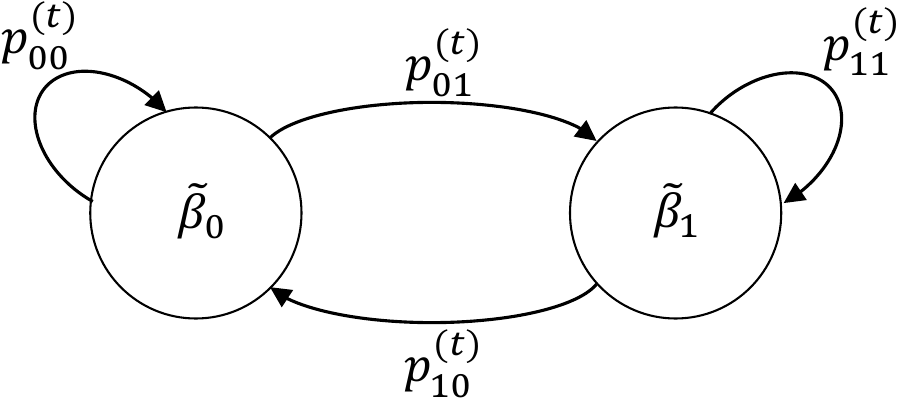}
		\vspace{-0.5em}
		\caption{Transition probability diagram for the \emph{channel state} at the beginning of decision cycle~$t$.} 
		\label{fig:mob_twoStateChannel}
		\vspace{-1em}
	\end{figure}

	In Fig.~\ref{fig:mob_twoStateChannel}, we show these states with their respective transition probabilities. For example, $p_{01}^{(t)}$ is the probability of the channel to transition from $\widetilde{\beta}_{0}$ to $\widetilde{\beta}_{1}$ at the \emph{beginning of decision cycle $t$}. Please note that $p_{11}^{(t)} + p_{10}^{(t)} = 1$ and $p_{00}^{(t)} + p_{01}^{(t)} = 1$, thus, to characterize the transition probabilities, we only need to define $p_{11}^{(t)}$ and $p_{01}^{(t)}$. Since these probabilities are a function of the serving distance, they are dependent on the decision cycle $t$ (as can be seen in Fig.~\ref{fig:mob_twoStateChannel}) which is not always the case for different POMDP problems.

	\subsubsection{Belief}
	Let us denote a specific value $\widetilde{\bf s}_i$ 
	for the state, then the belief $\omega_{i, \mathcal{B}_{\rm cand, \ell}}^{(t)}$ of the current state ${\bf s}^{(t)}$ to be equal to $\widetilde{\bf s}_i \in \mathcal{S}$ is written as
	\begin{align}\label{eq:belief_w}
		\omega_{i, \mathcal{B}_{\rm cand, \ell}}^{(t)} &= \mathbb{P}\left( {\bf s}^{(t)} = \widetilde{\bf s}_i |\ \mathcal{H}_{t-1}\right)
		\nonumber \\
		&=
		\prod_{b \in \mathcal{B}_{\rm cand, \ell}}
		\mathbb{P}\left( s_{bu}^{(t)} = \widetilde{s}_{i,bu}  |\ \mathcal{H}_{t-1}\right)
		,\quad i = 1,\dots, |\mathcal{S}|
	\end{align}
	We define the belief of state $s_{bu}^{(t)}$ being in $\widetilde{\beta}_{1}$ state as $\Upsilon_{bu}^{(t)}$ and that of being in $\widetilde{\beta}_{0}$ as $(1 - \Upsilon_{bu}^{(t)})$. Hence, the probability $\mathbb{P}\left( s_{bu}^{(t)} = \widetilde{s}_{i,bu}  |\ \mathcal{H}_{t-1}\right)$ in~\eqref{eq:belief_w} is defined as
	\vspace{-0.5em}
	\begin{align}
		\mathbb{P}\left( s_{bu}^{(t)} = \widetilde{s}_{i,bu}  |\ \mathcal{H}_{t-1}\right) =
		\begin{cases}
			\Upsilon_{bu}^{(t)},& \text{if}\ \widetilde{s}_{i,bu} = \widetilde{\beta}_{1} \\
			1 - \Upsilon_{bu}^{(t)},&  \text{if}\ \widetilde{s}_{i,bu} = \widetilde{\beta}_{0}
		\end{cases}
	\end{align}
	Then,
	%
	%
	%
	\begin{align}\label{eq:ProbGoodStateElement}
		&\Upsilon_{bu}^{(t)}
		\triangleq
		\mathbb{P}\left( s_{bu}^{(t)} = \widetilde{\beta}_{1}  |\ \mathcal{H}_{t-1}\right)
		=
		\nonumber \\
		&
		\begin{cases}
			p_{11, bu}^{(t)},& \text{if}\ a_{bu}^{(t-1)} = 1, s_{bu}^{(t-1)} = \widetilde{\beta}_{1}, \\
			p_{01, bu}^{(t)},&  \text{if}\ a_{bu}^{(t-1)} = 1, s_{bu}^{(t-1)} = \widetilde{\beta}_{0},\\
			\resizebox{0.48\linewidth}{!}
			{$
			\Upsilon_{bu}^{(t-1)} p_{11, bu}^{(t)}
			+ \left(1 - \Upsilon_{bu}^{(t-1)}\right) p_{01, bu}^{(t)}
			$}
			,&  \text{if}\ a_{bu}^{(t-1)} = 0
		\end{cases}
	\end{align}
	The first two cases in~\eqref{eq:ProbGoodStateElement} are used when the user is connected to AP $b$ at $(t-1)$. The third case is written in a probabilistic form because when $a_{bu}^{(t-1)} = 0$ the channel state $s_{bu}^{(t-1)}$ is~unobservable.
	
	For the initial belief ${\bm \omega}^{(0)}_{\mathcal{B}_{\rm cand, \ell}}$, we can either use a uniform distribution (commonly used in POMDP problems), or make use of the initial transition probabilities of the channels to define the elements $\{\omega_{i, \mathcal{B}_{\rm cand, \ell}}^{(0)} : i = 1,\dots, |\mathcal{S}|\}$ of the initial belief as 
	\begin{align}\label{eq:belief_w_initial}
		\omega_{i, \mathcal{B}_{\rm cand, \ell}}^{(0)} &=
		\mathbb{P}\left( {\bf s}^{(0)} = \widetilde{\bf s}_i |\ \mathcal{H}_{-1}\right)
		=
		\prod_{b \in \mathcal{B}_{\rm cand, \ell}}
		\mathbb{P}\left( s_{bu}^{(0)} = \widetilde{s}_{i,bu} \right)
	\end{align}
	with
	\begin{align}
		&\mathbb{P}\left( s_{bu}^{(0)} = \widetilde{s}_{i,bu} \right)
		\nonumber \\
		&
		=
		\begin{cases}
			1,& \text{if}\ b \in \mathcal{C}_u^{(0)}, \beta_{bu}^{(0)} > \beta_{\rm threshold}, \widetilde{s}_{i,bu} = \widetilde{\beta}_{1} \\
			1,& \text{if}\ b \in \mathcal{C}_u^{(0)}, \beta_{bu}^{(0)} \le \beta_{\rm threshold}, \widetilde{s}_{i,bu} = \widetilde{\beta}_{0} \\
			0,& \text{if}\ b \in \mathcal{C}_u^{(0)}, \beta_{bu}^{(0)} > \beta_{\rm threshold}, \widetilde{s}_{i,bu} = \widetilde{\beta}_{0} \\
			0,& \text{if}\ b \in \mathcal{C}_u^{(0)}, \beta_{bu}^{(0)} \le \beta_{\rm threshold}, \widetilde{s}_{i,bu} = \widetilde{\beta}_{1} \\
			\bar{p}_{1, bu}^{(0)},& \text{if}\ b \notin \mathcal{C}_u^{(0)}, \widetilde{s}_{i,bu} = \widetilde{\beta}_{1} \\
			\bar{p}_{0, bu}^{(0)},&  \text{if}\ b \notin \mathcal{C}_u^{(0)}, \widetilde{s}_{i,bu} = \widetilde{\beta}_{0}
		\end{cases}&&
	\end{align}
	where the terms $\bar{p}_{1, bu}^{(t)}$ and $\bar{p}_{0, bu}^{(t)}$ are the probabilities of initially observing $\widetilde{\beta}_{1}$ or $\widetilde{\beta}_{0}$, respectively; in such a case the channel state is not known. These probabilities are defined in~\eqref{eq:p1} and~\eqref{eq:p0}. The elements of the belief vector sums to one, i.e., $\sum_{i = 1}^{|\mathcal{S}|} \omega_{i, \mathcal{B}_{\rm cand, \ell}}^{(t)} = 1$.
	
	\subsubsection{Transition and Observation Probabilities}
	In addition to the belief vector, the elements of the transition probability matrix in~\eqref{eq:Transition_prob} are defined as
	\begin{align}\label{eq:Transition_prob_details}
		[{\bf P}_{\rm s, {\mathcal{B}_{\rm cand, \ell}}}^{(t)}]_{i,i'}
		=
		\prod_{b \in \mathcal{B}_{\rm cand, \ell}}
		\mathbb{P}( s_{bu}^{(t)} = \widetilde{s}_{i',bu} |\ s_{bu}^{(t-1)} = \widetilde{s}_{i,bu} )
	\end{align}
	with
	\begin{align}
		&\mathbb{P} \left( s_{bu}^{(t)} = \widetilde{s}_{i',bu} |\ s_{bu}^{(t-1)} = \widetilde{s}_{i,bu} \right)
		=
		\nonumber \\
		&
		\quad\quad\quad\quad
		\begin{cases}
			p_{11, bu}^{(t)},& \text{if}\ \widetilde{s}_{i',bu} = \widetilde{\beta}_{1}, \widetilde{s}_{i,bu} = \widetilde{\beta}_{1} \\
			p_{01, bu}^{(t)},& \text{if}\ \widetilde{s}_{i',bu} = \widetilde{\beta}_{1}, \widetilde{s}_{i,bu} = \widetilde{\beta}_{0}\\
			1 - p_{11, bu}^{(t)},& \text{if}\ \widetilde{s}_{i',bu} = \widetilde{\beta}_{0}, \widetilde{s}_{i,bu} = \widetilde{\beta}_{1}\\
			1 - p_{01, bu}^{(t)},& \text{if}\ \widetilde{s}_{i',bu} = \widetilde{\beta}_{0}, \widetilde{s}_{i,bu} = \widetilde{\beta}_{0}
		\end{cases}
	\end{align}
	
	Moreover, the observation probability is defined as follows.
	\begin{align}\label{eq:observ_prob}
		&\left[{\bf P}_{\rm o, \mathcal{B}_{\rm cand, \ell}}^{(t)}\right]_{i,l,j}
		=
		\mathbb{P}({\bf o}^{(t)} = \widetilde{\bf o}_l|\ {\bf s}^{(t)} = \widetilde{\bf s}_i, {\bf a}^{(t)} = \widetilde{\bf a}_j)
		\nonumber \\
		& \ 
		=
		\prod_{b \in \mathcal{B}_{\rm cand, \ell}} \mathbb{P}\left( {o}_{bu}^{(t)} = \widetilde{o}_{l,bu} |\  {\bf s}^{(t)} = \widetilde{\bf s}_i, {\bf a}^{(t)} = \widetilde{\bf a}_j\right)
		\nonumber \\
		& \ 
		=
		\Big(
		\prod_{b \in \mathcal{C}_u^{(t)}}
		\underbrace{
			\zeta_{bu, ilj}
		}_{\text{Observed channel state}}
		\Big)
		\prod_{b \in \mathcal{B}_{\rm cand, \ell} \backslash\mathcal{C}_u^{(t)}} 
		\underbrace{
			\bar{\zeta}_{bu, ilj}
		}_{\text{Non-observed channel state}}
	\end{align}
	where the observed and non-observed channel states in~\eqref{eq:observ_prob} are, respectively, defined as follows
	\begin{align}
		\zeta_{bu, ilj} & \triangleq
		\mathbb{P}\left( {o}_{bu}^{(t)} = \widetilde{s}_{l,bu} |\ {\bf s}^{(t)} = \widetilde{\bf s}_i, {\bf a}^{(t)} = \widetilde{\bf a}_j\right)
		\nonumber \\
		&
		=
		\begin{cases}
			1, & \text{if}\ l = i \\
			0, & \text{otherwise}
		\end{cases}	
		\\
		\bar{\zeta}_{bu, ilj} & \triangleq 
		\mathbb{P}\left( {o}_{bu}^{(t)} = \widetilde{o}_{l,bu} |\  {\bf s}^{(t)} = \widetilde{\bf s}_i, {\bf a}^{(t)} = \widetilde{\bf a}_j\right)
		\nonumber \\
		&
		=
		\begin{cases}
			\bar{p}_{1, bu}^{(t)}, & \text{if}\ \widetilde{o}_{l,bu} = \widetilde{\beta}_{1} \\
			\bar{p}_{0, bu}^{(t)}, & \text{if}\ \widetilde{o}_{l,bu} = \widetilde{\beta}_{0} 
		\end{cases}
		\label{eq:observProb_unknown}
	\end{align}
	where $\bar{p}_{1, bu}^{(t)}$ and $\bar{p}_{0, bu}^{(t)}$ are defined next in~\eqref{eq:p1} and~\eqref{eq:p0}.
	\begin{remark}
		If no information is known at all for the observation probability, a natural value for~\eqref{eq:observProb_unknown} is $0.5$. This means that the product term over $\mathcal{B}_{\rm cand, \ell}\backslash\mathcal{C}_u^{(t)}$ in~\eqref{eq:observ_prob} is $\frac{1}{2^{|\mathcal{B}_{\rm cand, \ell}| - |\mathcal{B}_{\rm con}| }}$. For example, when the user is connected to all the APs except for a single one, then we have a single state that is unknown (i.e., $|\mathcal{B}_{\rm cand, \ell}| - |\mathcal{B}_{\rm con}| = 1$), and this channel can be either $\widetilde{\beta}_{1}$ or $\widetilde{\beta}_{0}$ with equal probabilities of $0.5$.
	\end{remark}
	To define the terms $\bar{p}_{1, bu}^{(t)}$, $\bar{p}_{0, bu}^{(t)}$, $p_{11, bu}^{(t)}$ and $p_{01, bu}^{(t)}$, we need to first define a model that describes the spatial correlation for the LSF as the user moves throughout the network area. For a smooth flow for our exposition, we define this spatial correlation in Appendix~\ref{sec:SpatialCorrelationLSF}.
	
	\subsubsection{Channel State observation probabilities $\bar{p}_{1, bu}^{(t)}$ and $\bar{p}_{0, bu}^{(t)}$}
	The probability of observing $\widetilde{\beta}_{1}$ between AP $b$ and the typical user at an initial decision cycle of the POMDP formulation can be defined~as
	\begin{align}\label{eq:p1}
		\bar{p}_{1, bu}^{(t)} &
		\triangleq
		\mathbb{P}(s_{bu}^{(t)} = \widetilde{\beta}_{1} )
		\stackrel{(a)}
		{=}
		\mathbb{P}(\beta_{bu}^{(t)} > \beta_{\rm threshold} )
		\nonumber \\
		&
		\stackrel{(b)}
		{=}
		\mathbb{P}\left(\bar{\kappa}_{bu}^{(t)} > \frac{10}{\sigma_\mathrm{sh|dB}}\log_{10}\left(\frac{\beta_{\rm threshold}}{{\rm PL}^{(t)}}\right) \right)
		=
		{\rm Q}\left( \dot{k}^{(t)} \right)
		,
	\end{align}
	where $(a)$ follows from \eqref{eq:twoChannelState}, $(b)$ follows from the discussion in Appendix~\ref{sec:SpatialCorrelationLSF} for $t=0$ (check~\eqref{eq:shad_model}), ${\rm PL}^{(t)}$ is the path loss defined in~\eqref{eq:PL}, and  $\bar{\kappa}_{bu}^{(t)} \sim \mathcal{N}(0, 1)$, ${\rm Q}(\cdot)$ is the Q-function, and the argument $\dot{k}^{(t)} = \frac{10}{\sigma_\mathrm{sh|dB}}\log_{10}\left(\frac{\beta_{\rm threshold}}{{\rm PL}^{(t)}}\right)$. Furthermore, by definition, the probability of observing $\widetilde{\beta}_{0}$ between AP $b$ and the typical user~equals
	\begin{align}\label{eq:p0}
		\bar{p}_{0, bu}^{(t)} 
		\triangleq
		1 - \bar{p}_{1, bu}^{(t)}
	\end{align}

	\subsubsection{Channel State Transition probabilities $p_{11, bu}^{(t)}$ and $p_{01, bu}^{(t)}$}
	The transition probability when $s_{bu}^{(t-1)} = s_{bu}^{(t)} = \widetilde{\beta}_{1}$, for $t>0$, can be calculated as
	\begin{align}
		p_{11, bu}^{(t)}
		&\triangleq
		\mathbb{P}(s_{bu}^{(t)} = \widetilde{\beta}_{1} |\ s_{bu}^{(t-1)} = \widetilde{\beta}_{1} )
			\nonumber \\
			&
		\stackrel{(a)}
		{=}
		\mathbb{P}(\beta_{bu}^{(t)} > \beta_{\rm threshold} |\ \beta_{bu}^{(t-1)} > \beta_{\rm threshold} )
		\nonumber \\
		&=
		\mathbb{P}\left(\bar{\kappa}_{bu}^{(t)} > \frac{10}{\sigma_\mathrm{sh|dB}}\log_{10}\left(\frac{\beta_{\rm threshold}}{{\rm PL}^{(t)}}\right) 
		\right.
		\nonumber \\
		& \quad\quad\quad
		\left.
		|\  \bar{\kappa}_{bu}^{(t-1)} > \frac{10}{\sigma_\mathrm{sh|dB}}\log_{10}\left(\frac{\beta_{\rm threshold}}{{\rm PL}^{(t-1)}}\right) \right)
	\end{align}
	where, again, $(a)$ follows from \eqref{eq:twoChannelState}, $(b)$ follows from the discussion in Appendix~\ref{sec:SpatialCorrelationLSF} for $t>0$ (check~\eqref{eq:shad_model}), and $\bar{\kappa}_{bu}^{(t-1)}, \bar{\kappa}_{bu}^{(t)} \sim \mathcal{N}(0, 1)$. Let us denote the probability density function (PDF) of $\bar{\kappa}_{bu}^{(t-1)}$ and $\bar{\kappa}_{bu}^{(t)}$ as $f_{\bar{K}}(\cdot)$, and define the argument $\dot{k}^{(t-1)} = \frac{10}{\sigma_\mathrm{sh|dB}}\log_{10}\left(\frac{\beta_{\rm threshold}}{{\rm PL}^{(t-1)}}\right)$, then, we~have 
	\begin{align}\label{eq:p11}
		p_{11, bu}^{(t)}
		&
		\triangleq
		\mathbb{P}\left(\bar{\kappa}_{bu}^{(t)} > \dot{k}^{(t)} |\  \bar{\kappa}_{bu}^{(t-1)} > \dot{k}^{(t-1)} \right)
			\nonumber \\
			&
		=
		\frac{
			\mathbb{P}\left(\bar{\kappa}_{bu}^{(t)} > \dot{k}^{(t)} ,  \bar{\kappa}_{bu}^{(t-1)} > \dot{k}^{(t-1)} \right)
		}
		{
			\mathbb{P}\left( \bar{\kappa}_{bu}^{(t-1)} > \dot{k}^{(t-1)} \right)
		}
		\nonumber \\
		&=
		\frac{
			\int_{\dot{k}^{(t)}}^{\infty} \int_{\dot{k}^{(t-1)}}^{\infty}
			f_{\bar{\kappa}_{bu}^{(t-1)}, \bar{\kappa}_{bu}^{(t)}}\left(k_1, k_2 \right)
			\diff k_1 \diff k_2
		}
		{
			{\rm Q}\left( \dot{k}^{(t-1)} \right)
		}
	\end{align}
	where
	\begin{align}
	&f_{\bar{\kappa}_{bu}^{(t-1)}, \bar{\kappa}_{bu}^{(t)}}\left(k_1, k_2 \right) = \frac{1}{2\pi} (\det({\bm \Sigma}_{\bar{k}}))^{-1/2} 
	\nonumber \\
	& \quad \quad \quad
	\times
	\exp\left(-0.5 \times [k_1,\ k_2] \times {\bm \Sigma}_{\bar{k}}^{-1} \times  [k_1,\  k_2]^T\right),
	\end{align}
	is the PDF of the bivariate Gaussian distribution for $(\bar{\kappa}_{bu}^{(t-1)}, \bar{\kappa}_{bu}^{(t)})$, where ${\bm \Sigma}_{\bar{k}} \in \mathbb{R}^{2 \times 2}$ is the covariance matrix with $\left[{\bm \Sigma}_{\bar{k}}\right]_{11} = \left[{\bm \Sigma}_{\bar{k}}\right]_{22} = 1$ and $\left[{\bm \Sigma}_{\bar{k}}\right]_{12} = \left[{\bm \Sigma}_{\bar{k}}\right]_{21} = \iota + \left(1 - \iota\right) 2^{-\frac{\mathtt{v}_u \bar{\Delta} }{ d_\mathrm{decorr}}}$, which is obtained using the spatial correlation model defined in Appendix~\ref{sec:SpatialCorrelationLSF}. 


	Similarly, for $p_{01, bu}^{(t)}$, the transition probability when $s_{bu}^{(t-1)} = \widetilde{\beta}_{0}$ and $s_{bu}^{(t)} = \widetilde{\beta}_{1}$, for $t>0$, can be calculated as
	\begin{align}\label{eq:p01}
		p_{01, bu}^{(t)}
		&=
		\mathbb{P}(s_{bu}^{(t)} = \widetilde{\beta}_{1} |\ s_{bu}^{(t-1)} = \widetilde{\beta}_{0} )
			\nonumber \\
			&
		=
		\mathbb{P}(\beta_{bu}^{(t)} > \beta_{\rm threshold} |\ \beta_{bu}^{(t-1)} \le \beta_{\rm threshold} )
		\nonumber \\
		&=
		\frac{
			\int_{\dot{k}^{(t)}}^{\infty} \int_{-\infty}^{\dot{k}^{(t-1)}}
			f_{\bar{\kappa}_{bu}^{(t-1)}, \bar{\kappa}_{bu}^{(t)}}\left(k_1, k_2 \right)
			\diff k_1 \diff k_2
		}
		{
			1 - {\rm Q}\left( \dot{k}^{(t-1)} \right)
		}
	\end{align}
	
	It can be clearly seen from~\eqref{eq:p11} and~\eqref{eq:p01} that the transition probabilities of the channel states are non-stationary. This means that the HO policy, obtained from solving the POMDP, can expire and some counter measures are needed.

	\setlength{\textfloatsep}{0pt}
	\begin{algorithm}[t]
		\footnotesize
		\SetAlgoLined
		\SetInd{0.1em}{1em}
		\caption{ HO Policy using POMDP: $\mathtt{POMDP}$ } 
		\label{algorithm:POMDP_divide}
		\textbf{Input:} $u$, $\mathcal{B}$, $\mathcal{B}_{\rm base}$, $B_{\rm con}$, $T_{\rm H}$, $\{{\beta}_{bu} : b \in \mathcal{B}_{\rm base} \}$ \label{step:Algo_input}\\
		\textbf{Output:} HO policy $\Pi^\star$ and candidate APs $\mathcal{B}_{\rm cand}^\star$ \label{step:Algo_output}\\
		Construct $\mathcal{B}_{\rm others} = \mathcal{B} \backslash \mathcal{B}_{\rm base}$.\label{step:Alog_othersPool}\\
		Initialize ${\rm loop}_{\rm max} = |\mathcal{B}| - B_{\rm con}$ and $\ell = 1$\label{step:Alog_init_POMDPSize}. \Comment{\texttt{Number of POMDP sub-problems}}\\
		\While{$\ell \le {\rm loop}_{\rm max}$}{\label{step:Algo_terminate} 
			Construct a pool of candidate APs $\mathcal{B}_{\rm cand, \ell} = \{\mathcal{B}_{\rm base} \cup b'\}$, where $b' = \mathcal{B}_{\rm others}(\ell)$.\label{step:Algo_candDUs}
			\Comment{\texttt{Divide-and-conquer}}
			\\
			Construct POMDP sub-problem $\bar{\mathcal{P}}_\ell = \mathcal{P}(\mathcal{B}_{\rm cand, \ell}, T_{\rm H})$ with ${\bf P}_{\rm s, \mathcal{B}_{\rm cand, \ell}}^{(t)}$, ${\bf P}_{\rm o, \mathcal{B}_{\rm cand, \ell}}^{(t)}(\widetilde{\bf o}_l)$, ${\bm \omega}^{(0)}_{\mathcal{B}_{\rm cand, \ell}}$, and $R_{\mathcal{B}_{\rm cand, \ell}}\left({\bf s}^{(t)}, {\bf a}^{(t)}\right)$ defined using~\eqref{eq:Transition_prob_details}, \eqref{eq:observ_prob}, and~\eqref{eq:belief_w_initial}, and~\eqref{eq:reward_SingleU} respectively.\label{step:Algo_POMDP}\\
			Solve $\bar{\mathcal{P}}_\ell$ to obtain $\Pi_\ell = \{\pi_{1}, \dots,  \pi_{T_{\rm H}}\}$ for a finite $T_{\rm H}$.
			\label{step:Algo_POMDPSolve}
			\Comment{\texttt{Using the Finite Grid algorithm}}\\
			Obtain total expected reward 
			from the solved $\bar{\mathcal{P}}_\ell$.\label{step:Algo_POMDPReward}\\
			$\ell = \ell + 1$ \Comment{\texttt{Next POMDP sub-problem}}\\
		}
		Select ${\rm loopOpt} = \underset{\ell}{\arg\max}\ \tilde{R}_\ell$.\label{step:Algo_POMDPIndexOfOpt} \Comment{\texttt{Select best POMDP sub-problem}}\\
		Obtain $\Pi^\star = \Pi_{\rm loopOpt}$ and $\mathcal{B}_{\rm cand}^\star = \mathcal{B}_{\rm cand, loopOpt}$\label{step:Algo_optPolicy}. \Comment{\texttt{Return policy \& APs' candidate pool}}\\
	\end{algorithm}

	\vspace{-1em}
	\subsection{POMDP Sub-problems and Time Horizon $T_{\rm H}$}\label{sec:POMDP_subproblems}
	\newcommand{\varComplexityOne}{The computational complexity of our problem is still a main issue as discussed previously, because the size of the state space is $2^{|\mathcal{B}|}$, which prevents us from solving the problem for even a reasonable value of $|\mathcal{B}|$, the number of APs.}
	
	\newcommand{\varComplexityTwo}{In this section, we propose to follow a divide-and-conquer approach to solve the POMDP problem. We divide the POMDP problem~$\mathcal{P}( \mathcal{B}, T_{\rm H})$ in~\eqref{eq:POMDP_model} into many POMDP sub-problems, each defined as $\bar{\mathcal{P}}_{\rm \ell} = \mathcal{P}( \mathcal{B}_{\rm cand, \ell}, T_{\rm H})$, where $\mathcal{B}_{\rm cand, \ell} \ll \mathcal{B}$. The advantages are two-fold; the first is that complexity is decreased when dealing with many small POMDP sub-problems compared to dealing with a single POMDP problem, and the second is that the complexity of a each POMDP sub-problem is independent of the number of APs found in the network. Taken together, our approach provides a scalable solution.}
	
	\varComplexityOne
	
	\varComplexityTwo
	
	However, the usage of POMDP sub-problems necessitates that we enforce an expiration time duration for the derived policy that uses a specific candidate pool of APs. After the expiry of the policy we need to re-derive a new policy that possibly uses a new candidate pool of APs. This expiration time window can be set using a finite time horizon $T_{\rm H}$ for the POMDP.

	We propose to solve our POMDP formulation using the procedure called~$\mathtt{POMDP}$ defined in Algorithm~\ref{algorithm:POMDP_divide}. This procedure formulates the POMDP framework and solves it to derive a HO policy in UC-mMIMO networks. Step~\ref{step:Algo_input} specifies the inputs used by Algorithm~\ref{algorithm:POMDP_divide}. The set $\mathcal{B}_{\rm base}$ contains the APs serving the user at the time of constructing the POMDP formulation. As can be seen, the LSF toward only these APs are known $\{{\beta}_{bu} : b \in \mathcal{B}_{\rm base} \}$. The term $B_{\rm con}$ is the number of APs the user will be connecting to, and $T_{\rm H}$ is the time horizon for POMDP.
	
	The outputs for the algorithm, shown in Step~\ref{step:Algo_output}, are the HO policy $\Pi^\star$ and the candidate AP pool $\mathcal{B}_{\rm cand}^\star$ used by $\Pi^\star$. The policy $\Pi^\star$ allows us to obtain the best action at any decision cycle within $T_{\rm H}$, which means that it determines which APs, from $\mathcal{B}_{\rm cand}^\star$, the user will connect to based on the current network belief ${\bm \omega}^{(t)}_{\mathcal{B}_{\rm cand}^\star}$. 
	
	Step~\ref{step:Alog_othersPool} defines the pool $\mathcal{B}_{\rm others}$ of APs to which the user is not currently connected to. We note that because the order of the elements in the sets $\mathcal{B}_{\rm others}$ is relevant, we can refer to $\mathcal{B}_{\rm others}$ as a sequence rather than a set. Step~\ref{step:Alog_init_POMDPSize} performs some initialization such as 
	setting the number of POMDP sub-problems ${\rm loop}_{\rm max}$ that will be considered later, and initializing a counter ${\rm \ell}$. Each loop $\ell$ in Step~\ref{step:Algo_terminate} corresponds to constructing and solving a single POMDP sub-problem that uses a specific pool of APs $\mathcal{B}_{\rm cand, \ell}$. Step~\ref{step:Algo_candDUs} constructs a candidate AP pool $\mathcal{B}_{\rm cand, \ell}$ of size $(B_{\rm con} + 1 )$ APs. Step~\ref{step:Algo_POMDP} constructs a POMDP model using $\mathcal{B}_{\rm cand, \ell}$.
	
	\begin{algorithm}[t]
		\footnotesize
		\SetAlgoLined
		\SetInd{0.1em}{1em}
		\caption{Apply HO Policy}
		\label{algorithm:Apply_Policy}
		\textbf{Input:} $u$, $\mathcal{B}$, $B_{\rm con}$, 
		$T_{\rm H}$, $\mathcal{C}_u^{(0)}$\\
		\textbf{Output:} User association $\mathcal{C}_u^{(t)}$ during mobility\\
		Set $\bar{\mathfrak{t}} = 1$ and $t = 1$. \\
		\While{TRUE}{
			$[\Pi^\star, \mathcal{B}_{\rm cand}^\star] = \mathtt{POMDP}\big(
			u$, $\mathcal{B}$, $\mathcal{B}_{\rm base} = \mathcal{C}_u^{(t-1)}$, $B_{\rm con}$, $T_{\rm H}$, $\{{\beta}_{bu}^{(t-1)} : b \in \mathcal{B}_{\rm base} \} \big)$
			\Comment{\texttt{Policy \& candidate pool}}
			\\
			\While{$t \le ((\bar{\mathfrak{t}} - 1) T_{\rm H} + T_{\rm H})$}{\label{step:Algo_ApplyPolicy}
				Calculate belief vector ${\bm \omega}^{(t)}_{\mathcal{B}_{\rm cand}^\star} = [\omega_{1, \mathcal{B}_{\rm cand}^\star}^{(t)}, \dots, \omega_{|\mathcal{S}|, \mathcal{B}_{\rm cand}^\star}^{(t)}]$ using~\eqref{eq:belief_w}.\\
				Choose ${\bf a}^{(t)} = \Pi^\star({\bm \omega}^{(t)}_{\mathcal{B}_{\rm cand}^\star})$ 
				\Comment{\texttt{Obtain action from policy}}\\
				Set $\mathcal{C}_u^{(t)} =
				\{ \left( {\rm diag} \left( {\bf a}^{(t)} [b_1 b_2 \dots b_{B_{\rm con}+1}] \right) \right)_{\ne 0}: [b_1 b_2 \dots b_{B_{\rm con}+1}] = \mathcal{B}_{\rm cand}^\star \}$\label{step:Algo_ApplyPolicy_ServingC}\\
				$t = t + 1$ 				\Comment{\texttt{Next decision cycle}}
			}
			$\bar{\mathfrak{t}} = \bar{\mathfrak{t}} + 1$
			\Comment{\texttt{New HO policy}}
		}
	\end{algorithm}

	Step~\ref{step:Algo_POMDPSolve} solves the POMDP model using the Finite Grid Algorithm~\cite{pomdpR}. This algorithm implements a variation of the point-based value iteration (PBVI)~\cite{pineau2003point} that allows for the solving of large POMDP problems. PBVI approximates an exact value iteration solution by selecting a small set of belief points and then it tracks the value for those points only. This approach helps in enhancing the scalability of the POMDP compared to the conventional value iteration algorithm~\cite{sondik1971optimal}. The Finite Grid Algorithm is well-known for solving POMDP problems, and it can be accessed through the R package {$\mathtt{pomdp}$}~\cite{pomdpR}. Since this algorithm is not part of our contributions, and due to the limited space of the paper, we refer the readers to~\cite{pineau2003point} for the details of the algorithm.

	Step~\ref{step:Algo_POMDPReward} obtains the total expected reward of the optimal solution obtained from the solver. This value is returned by default by the POMDP solver~\cite{pomdpR}. Step~\ref{step:Algo_POMDPIndexOfOpt} obtains the index of the POMDP sub-problem which produced the best total expected reward. Finally, Step~\ref{step:Algo_optPolicy} chooses the best policy as $\Pi^\star = \Pi_{\rm loopOpt}$ which corresponds for the AP candidate pool $\mathcal{B}_{\rm cand}^\star = \mathcal{B}_{\rm cand, loopOpt}$.
	
	Algorithm~\ref{algorithm:Apply_Policy} shows how to run the procedure $\mathtt{POMDP}$ (Algorithm~\ref{algorithm:POMDP_divide}) using a normal operation. 
	Algorithm~\ref{algorithm:Apply_Policy} is self-explanatory. It handles the call procedure of Algorithm~\ref{algorithm:POMDP_divide}, and the construction of the serving set $\mathcal{C}_u^{(t)}$ of the user using the actions obtained from the POMDP policy. Step~\ref{step:Algo_ApplyPolicy} marks the expiration of a HO policy and the start of a new one (here $T_{\rm H}$ is the time horizon), $t$ is the index of decision cycle and $\bar{\mathfrak{t}}$ is the index of the HO policy. We note that the operator $({\bf x})_{\ne 0}$ in Step~\ref{step:Algo_ApplyPolicy_ServingC} in Algorithm~\ref{algorithm:POMDP_divide} selects the nonzero entries in vector ${\bf x}$. In the next section, we will use this algorithm to produce different behaviors, such as minimizing the number of HOs.

	\begin{algorithm}[t]
		\footnotesize
		\SetAlgoLined
		\SetInd{0.1em}{1em}
		\caption{Apply HO Policy While Controlling the number of HOs}
		\label{algorithm:DecreaseHO}
		\textbf{Input:} $u$, $\mathcal{B}$, $B_{\rm con}$, $\mathcal{C}_u^{(0)}$, $R_u^{(\rm lb), (0)}$\\
		\textbf{Output:} User association $\mathcal{C}_u^{(t)}$ during mobility\\
		Set $t = 1$, $T_{\rm H} = 10$, and $\mathcal{C}_{\rm potential} = \mathcal{C}_u^{(0)}$.\label{step:Algo_ExpWindow}\\
		\While{TRUE}{
			$[\Pi^\star, \mathcal{B}_{\rm cand}^\star] = \mathtt{POMDP}\big(
			u$, $\mathcal{B}$, $\mathcal{B}_{\rm base} = \mathcal{C}_{\rm potential}$, $B_{\rm con}$, $T_{\rm H}$, $\{{\beta}_{bu}^{(t-1)} : b \in \mathcal{C}_u^{(t-1)} \} \big)$ \label{step:DecreaseHO_CallPOMDP} \\
			Calculate belief vector ${\bm \omega}^{(t)}_{\mathcal{B}_{\rm cand}^\star} = [\omega_{1, \mathcal{B}_{\rm cand}^\star}^{(t)}, \dots, \omega_{|\mathcal{S}|, \mathcal{B}_{\rm cand}^\star}^{(t)}]$ using~\eqref{eq:belief_w}. \label{step:DecreaseHO_belief}\\
			Choose ${\bf a}^{(t)} = \Pi^\star({\bm \omega}^{(t)}_{\mathcal{B}_{\rm cand}^\star})$\label{step:DecreaseHO_action} \Comment{\texttt{Obtain action from policy}}\\
			Choose $\mathcal{C}_{\rm potential} =
			\Big\{ \left( {\rm diag} \left( {\bf a}^{(t)} [b_1 b_2 \dots b_{B_{\rm con}+1}] \right) \right)_{\ne 0}: [b_1 b_2 \dots b_{B_{\rm con}+1}] = \mathcal{B}_{\rm cand}^\star \Big\} $\label{step:DecreaseHO_cPotential}\\
			\eIf{ $R_u^{(\rm lb), (t-1)} \ge R_{\rm threshold}$\label{step:DecreaseHO_applyCStart}}
			{
				Choose $\mathcal{C}_u^{(t)} = \mathcal{C}_u^{(t-1)}$ \Comment{\texttt{No HOs}}\\
			}{
				Choose $\mathcal{C}_u^{(t)} = \mathcal{C}_{\rm potential}$
				\Comment{\texttt{Obtain serving set using obtained action}}\\
			}\label{step:DecreaseHO_applyCEnd}
			Calculate $R_u^{(\rm lb), (t)}$ in~\eqref{eq:rate_LB} at decision cycle $t$.\\
			$t = t + 1$ \Comment{\texttt{Next decision cycle}}
		}
	\end{algorithm}

	\begin{table*}[t]
		\scriptsize
		\centering
		\begin{tabular}{|p{0.08\linewidth}|p{0.13\linewidth}|p{0.14\linewidth}||p{0.08\linewidth}|p{0.15\linewidth}|p{0.24\linewidth}|}
			\hline
			\hline
			\multicolumn{1}{|l|}{ \textit{\textbf{Description}}} & \multicolumn{1}{l|}{ \textit{\textbf{Parameter}}} & \multicolumn{1}{l||}{\textit{\textbf{Value}}}& \multicolumn{1}{l|}{ \textit{\textbf{Description}}} & \multicolumn{1}{l|}{ \textit{\textbf{Parameter}}} & \multicolumn{1}{l|}{\textit{\textbf{Value}}}\\
			\hline
			Network & $|\mathcal{B}|$, $M$, $\mathtt{v}_u$ & $125$, $8$, $10~{\rm m/s}$ &
			POMDP & 
				$\gamma$, $\beta_{\rm threshold}$, $\widetilde{\beta}_{1}$, $\widetilde{\beta}_{0}$, $B_{\rm con}$, 
				$\bar{\Delta}$
			& 
				\pbox{20cm}{$0.95$, ${\rm PL}(\bar{d} = 150~{\rm m})$,} 
				\pbox{10cm}{${\rm PL}(\bar{d} = 50~{\rm m})$,${\rm PL}(\bar{d} = 200~{\rm m})$},
				$5$, 
				$1~{\rm s}$
			\\
			\hline		
			\pbox{10cm}{Power, pilots} & $p^{({\rm d})}$, $p^{({\rm u})}$, $\tau_{\rm c}$, $\tau_{\rm p}$  & $30~{\rm dBm}$, $20~{\rm dBm}$, $200$, $16$ &
			Channel aging & ${\rm \lambda_0}$, $c_0$, $T_{\rm s}$ & $(3\times 10^8)/{\rm c}_0$, $1.8~{\rm GHz}$, $66.7~\mu{\rm s}$
			\\
			\hline
			Path loss \& fading &
			$d_0$, $\alpha_{\rm pl}$, $d_{\rm h}$, $\sigma_\mathrm{sh|dB}$, $d_\mathrm{decorr}$, $\iota$
			& $1.1~{\rm m}$, $3.8$, $13.5~{\rm m}$, $6~{\rm dB}$, $100~{\rm m}$, $0.5$ &
			Noise & spectral density $S_z$, noise figure $F_z$, BW & \pbox{10cm}{$-174~{\rm dBm/ Hz}$, $8~{\rm dB}$,} $20~{\rm MHz}$\\
			\hline
			\hline
		\end{tabular}
		\vspace{-0.5em}
		\caption{Simulation parameters.}
		\label{table:sim_parameters}   
		\vspace{-1.8em}
	\end{table*}

	\vspace{-1em}
	\section{Decreasing the Number of HOs}\label{sec:DecreasingNofHO}
	\vspace{-0.3em}
	The importance of using POMDP to derive a HO policy is that we do not need full observability of all the channel states, and we can consider future rewards and temporal evolution of the modeled process. Moreover, a POMDP can provide feedback control. Hence, as a concept, the feedback control can be tuned to perform any functionality or to prioritize any needed metric.

	With user-centric clustering and the absence of cells on the access channel, a moving user may experience a lot of HOs that may degrade the performance due to HO overhead. Hence, controlling the number of HOs is an important functionality, especially with a dense deployment of the APs. Hence, we set our (\emph{noncausal}) objective function and constraints as follows
	\begin{subequations}\label{eq:objMinHO}
		\begin{align}
			\min_{ \{\mathcal{C}_u^{(t)}: \forall t \} }
			\quad
			&
			\sum_{t = 1}^{\infty} N_{\rm HO}^{(t)}
			\label{eq:objMinHO_obj}
			\\
			\text{s.t.}
			\quad
			&
			R_u^{(\rm lb), (t)} \ge R_{\rm threshold}, \quad \forall t 
			\label{eq:objMinHO_cons_rate}
			\\
			&
			N_{\rm HO}^{(t)} = |\mathcal{C}_u^{(t)} - \mathcal{C}_u^{(t-1)}|, \quad \forall t 
			\label{eq:objMinHO_N}
		\end{align}
\end{subequations}
where $ \left( \mathcal{C}_u^{(t)} - \mathcal{C}_u^{(t-1)} \right)$ returns the elements of $ \mathcal{C}_u^{(t)}$ that are not found in $\mathcal{C}_u^{(t-1)}$,~\eqref{eq:objMinHO_cons_rate} constraints the spectral efficiency~\eqref{eq:rate_LB} to be above a chosen threshold $R_{\rm threshold}$, and constraint~\eqref{eq:objMinHO_N} defines the number of HOs.

Our problem is a Markov process, hence the formulation in~\eqref{eq:objMinHO} requires control over the temporal space $\forall t$. Thus, the problem cannot be approached using a conventional mathematical optimization technique, and hence we need to use our divide-and-conquer approach proposed in Algorithm~\ref{algorithm:POMDP_divide}. Based on this, within each time horizon $T_{\rm H}$, we propose employing the best policy $\Pi^\star$ obtained from Algorithm~\ref{algorithm:POMDP_divide} to maximize the reward of the POMDP which is a proxy for the spectral efficiency. We then use this policy only when the constraint in~\eqref{eq:objMinHO_cons_rate} is not satisfied. Otherwise, the serving set of the user is kept the same, which means we do not have HOs and the number of HOs is minimized. This approach is suboptimal, however, it allows to employ the POMDP to execute HOs only when needed. The solution steps are detailed in Algorithm~\ref{algorithm:DecreaseHO}.


	In Algorithm~\ref{algorithm:DecreaseHO}, we present a method to control the number of HOs and hence obtain the behavior described by~\eqref{eq:objMinHO}. In Step~\ref{step:Algo_ExpWindow} of this algorithm, we set $T_{\rm H}$ and define the potential serving set $\mathcal{C}_{\rm potential} = \mathcal{C}_u^{(0)}$, this cluster will be used as $\mathcal{B}_{\rm base}$ when calling Algorithm~\ref{algorithm:POMDP_divide}. A key difference between Algorithm~\ref{algorithm:DecreaseHO} and Algorithm~\ref{algorithm:Apply_Policy} is that $\mathcal{B}_{\rm base}$ does not always use $\mathcal{C}_u^{(t-1)}$, which changes the observability of the states because we can only observe the LSF for the APs that the user is currently connected to ($\{{\beta}_{bu}^{(t-1)} : b \in \mathcal{C}_u^{(t-1)} \}$). As a general rule, as $\mathcal{C}_{\rm potential}$ becomes more different from $\mathcal{C}_u^{(t-1)}$, the observability of the channel states reduces.
	
	Step~\ref{step:DecreaseHO_CallPOMDP} calls $\mathtt{POMDP}$ using $\mathcal{B}_{\rm base} = \mathcal{C}_{\rm potential}$. Steps~\ref{step:DecreaseHO_belief}, \ref{step:DecreaseHO_action}, and~\ref{step:DecreaseHO_cPotential} calculate the current belief, obtain the optimal action according to the derived policy, and construct $\mathcal{C}_{\rm potential}$ used in the POMDP formulations, respectively. Steps~\ref{step:DecreaseHO_applyCStart} till~\ref{step:DecreaseHO_applyCEnd} decide when to keep the serving set $\mathcal{C}_u^{(t)}$ the same (no HO), and when to change $\mathcal{C}_u^{(t)}$ by setting it to $\mathcal{C}_{\rm potential}$ obtained from POMDP. The condition to apply HO is based on having a spectral efficiency less than $R_{\rm threshold}$. The threshold $R_{\rm threshold}$ can be chosen based on the needed requirements, it also can be SINR or SNR-based.
	
	\setlength{\textfloatsep}{30pt}

	\section{Simulation Results}\label{sec:Results}
	We simulate networks of $|\mathcal{B}| = 125$~APs uniformly distributed in a $(1\times1)~{\rm km}^2$ network area. The typical user is initially located around the network center and moves on a straight line in a trip of distance $1000~{\rm m}$ at a speed of $10~{\rm m/s}$ ($36~{\rm km/h}$) through movements with a step size of duration $\bar{\Delta} = 1~{\rm s}$ (distance moved is $10~{\rm m}$). To study the performance of our approach, we apply network wrap around whenever the user reaches $200~{\rm m}$ from the network boundary, which emulates a mobile network with infinite area. The typical user and the APs are assumed to be found at heights~$1.5~{\rm m}$ and~$15~{\rm m}$, respectively, which enforces a minimum separation distance of $d_{\rm h} = 13.5~{\rm m}$ even when the user moves. This is achieved through our use of $d_{\rm h}$ in the path loss in~\eqref{eq:PL}. In Table~\ref{table:sim_parameters}, we summarize the remaining network parameters.

	\begin{figure}[t]
			\centering
			\includegraphics[width=1\columnwidth]{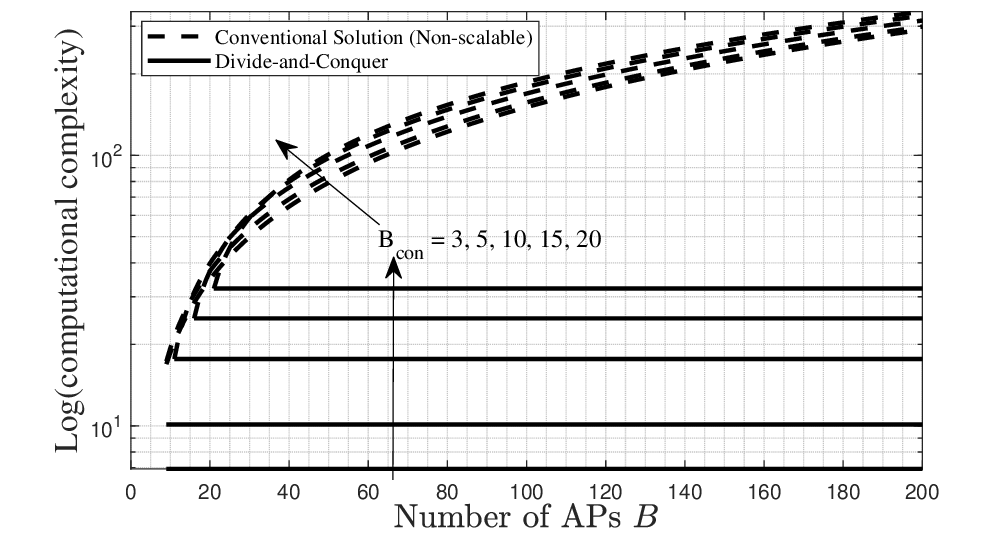}
			\vspace{-1.5em}
			\caption{Complexity of our solution compared to the conventional approach.} 
			\label{fig:complexity}
	\end{figure}
	
	\begin{table}[t]
		\scriptsize
		\centering
		\begin{tabular}{|p{0.22\linewidth}|p{0.13\linewidth}|p{0.2\linewidth}|p{0.24\linewidth}|}
			\hline
			\hline
			\multicolumn{1}{|l|}{ \textit{\textbf{Parameters}}} & \multicolumn{1}{l|}{ \textit{\textbf{LSF-based}}} &
			\textit{\textbf{Conventional Solution}}
			&
				\textit{\textbf{Divide-and-Conquer}}
			\\
			\hline
			Type & Passive
			& Active & Active
			\\
			\hline
			Observability Requirements & Fully & Partially & Partially
			\\
			\hline
			Future Rewards & No & Yes & Yes
			\\
			\hline
			POMDPs & -----
			& 1 & $(B - B_{\rm con})$
			\\
			\hline
			Consider temporal correlation & No & Yes & Yes
			\\
			\hline
			Complexity & $\mathcal{O}(B)$ & $\mathcal{O}\left(2^{2B} \right)$ & $\mathcal{O}\left(2^{2(B_{\rm con}+1)} \right)$
			\\
			\hline
			\hline
		\end{tabular}
		\vspace{-0.5em}
		\caption{Comparison between different methods.}
		\vspace{-3.5em}
		\label{table:complexity}
	\end{table}

	
	
	\begin{figure*}[t]
		\centering
		\begin{subfigure}[t]{0.48\textwidth}
			\centering			\includegraphics[width=1\textwidth]{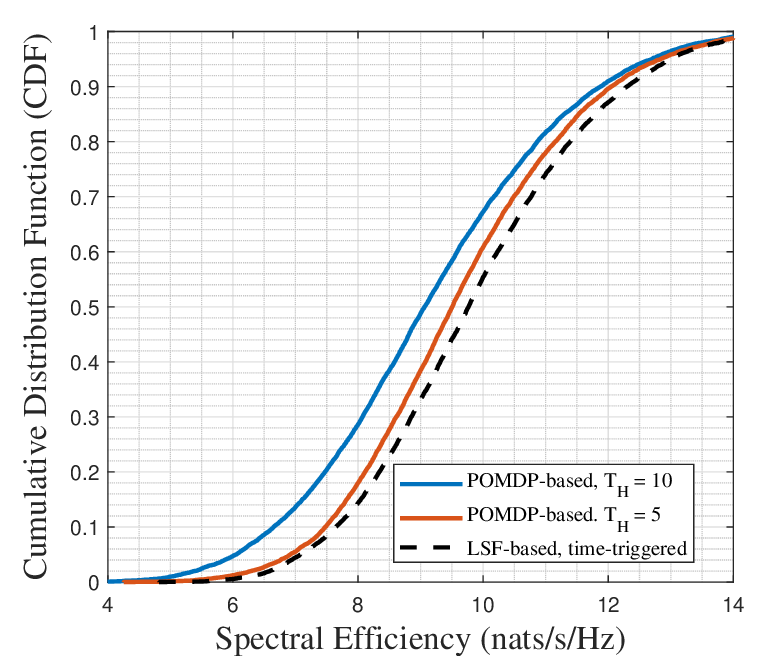}
			\vspace{-0.5em}
			\caption{Cumulative density function of spectral efficiency of typical user.}
			\label{fig:CDF}
		\end{subfigure}
		\quad\quad
		\begin{subfigure}[t]{0.44\textwidth}
			\centering			\includegraphics[width=1\textwidth]{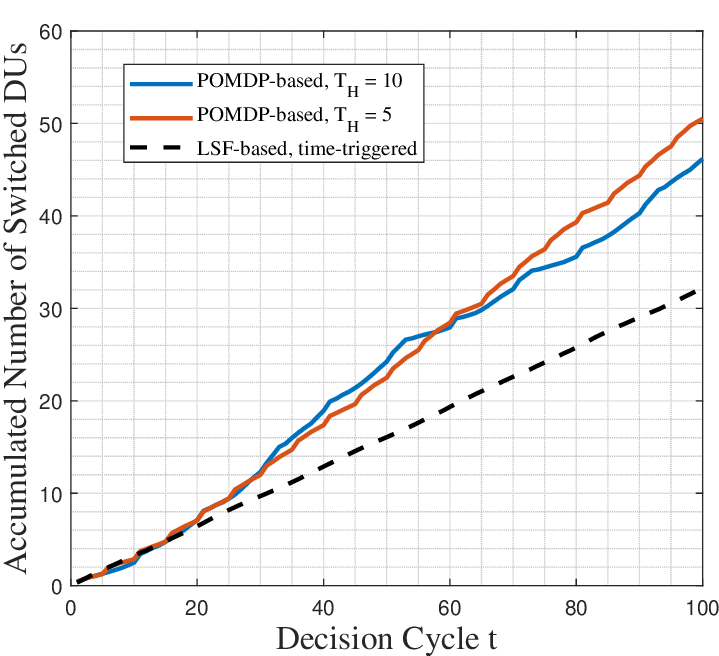}
			\vspace{-0.5em}
			\caption{Accumulative switched number of APs. 
			}
			\label{fig:avgSwitchedDUs_cummulative}
		\end{subfigure}
		\caption{Performance using Algorithm~\ref{algorithm:Apply_Policy} with different time horizon $T_{\rm H}$.}
		\label{fig:ResultSet1}
	\end{figure*}

	\label{page:varComplexityDiscussionOne}\newcommand{\varComplexityDiscussionOne}{
	The computational complexity of solving a POMDP sub-problem is $\mathcal{O}\left( |A| |\mathcal{S}|^2 \right)$ per iteration using the Value Iteration algorithm~\cite{condon1992complexity}, where we need to solve $|\mathcal{B}| - B_{\rm con}$ POMDP sub-problems. With a number of states of $|\mathcal{S}| = 2^{B_{\rm con}+1}$, with $(B_{\rm con} + 1)$ being the size of each POMDP sub-problem, we obtain a computational complexity of 
	$\mathcal{O}\left(2^{2 (B_{\rm con}+1) } \right)$. 
	This is much smaller than solving a single POMDP problem which has a computational complexity of 
	$\mathcal{O}\left( 2^{2 |\mathcal{B}|} \right)$, for a conventional solution. We note that a conventional approach suffers from a large state space for a large network size, which results in huge transition and observation probability matrices. 
	Hence, our solution bypasses this significant limitation.}
	
	\newcommand{\varComplexityDiscussionTwo}{In Fig.~\ref{fig:complexity}, we plot the computational complexity of the conventional approach and our divide-and-conquer approach, which shows the large advantage in using our solution. The price to be paid by our solution is that we need to solve $B - B_{\rm con}$ sub-problems compared to solving one problem for the conventional approach. However, the sub-problems can be defined and solved independently, which allows us to make use of prarallel computing to solve many sub-problems simultaneously. Moreover, our solution can be defined for \textit{any network size}, while the conventional solution can only be defined for toy examples (Fig.~\ref{fig:complexity}).}

	\newcommand{\varComplexityDiscussionThree}{In Table~\ref{table:complexity}, we show the different considerations for different solutions to manage HOs. The LSF-based approach has many drawbacks which include being a passive approach that cannot consider different objectives (rewards), temporal correlation of the system parameters, and future rewards. Moreover, it requires full observability of the system states to make HO decisions, which is essentially impossible to obtain in an actual deployment. On the other hand, the conventional POMDP approach cannot be defined for actual problems due to the high complexity and the curse of dimensionalty. This suggests that our divide-and-conquer approach is the most suitable solution that provides a tradeoff between complexity and optimality.}

	\varComplexityDiscussionOne
	
	\varComplexityDiscussionTwo
	
	\varComplexityDiscussionThree

	To benchmark our solution, we compare it to the ideal case with a LSF-based HO. In this scheme, after each movement (time-triggered), the user connects to the $B_{\rm con}$~APs providing the best LSF. The LSF-based HO represents the best HO decision for the user provided that frequent HOs do not incur any overhead on performance (ideal case, not found in actual implementation). Moreover, the LSF-based HO requires the knowledge of the LSF between the user and \emph{all} the APs in the network at each user movement, which is disadvantageous. Moreover, the LSF-based HO is a passive scheme that do not provide control based on possible future events or rewards.

	In Figs.~\ref{fig:ResultSet1}(\subref{fig:CDF}) and and~\ref{fig:ResultSet1}(\subref{fig:avgSwitchedDUs_cummulative}), we plot the cumulative density function (CDF) of the spectral efficiency and the accumulated number of switched (added) APs, respectively. These figures demonstrate the performance of Algorithm~\ref{algorithm:Apply_Policy}. Results show that increasing the time horizon $T_{\rm H}$ degrades the performance because the candidate set of APs that are used for the HO policy can become very far from the moving user, and the policy for these APs is still used to manage HOs. On the other hand, decreasing $T_{\rm H}$ provides a performance close to the LSF-based HO despite the fact that the POMDP solution assumes that the LSF for unconnected APs is not known (partial observation for state). This reduction in $T_{\rm H}$, however, increases the number of HOs.

	In Fig.~\ref{fig:ResultSet1}(\subref{fig:avgSwitchedDUs_cummulative}), we show the cumulative average number of switched APs as a function of decision cycles. These plots represent the number of APs added to the serving set of the user (the same number will be removed because $B_{\rm con}$ is fixed), i.e., this number represents the number of HOs. In general, Algorithm~\ref{algorithm:Apply_Policy} seems to present a higher number of HOs than LSF-based HO, however, this number seems to decrease as we increase $T_{\rm H}$. This result shows the importance of Algorithm~\ref{algorithm:DecreaseHO} used to decrease the number of HOs, with results presented next.

	To test Algorithm~\ref{algorithm:DecreaseHO}, we add another benchmark scheme. This scheme, denoted as ``LSF-based, threshold-triggered'', initiates HO decisions when the data rate drops below $R_{\rm threshold}$, but still the user connects to $B_{\rm con}$~APs providing the best LSF among all the APs in the network. Hence, the LSF between the user and all the APs is still needed.

\begin{figure*}[t]
		\centering
		\begin{subfigure}[t]{0.315\textwidth}
			\centering			\includegraphics[width=1\textwidth]{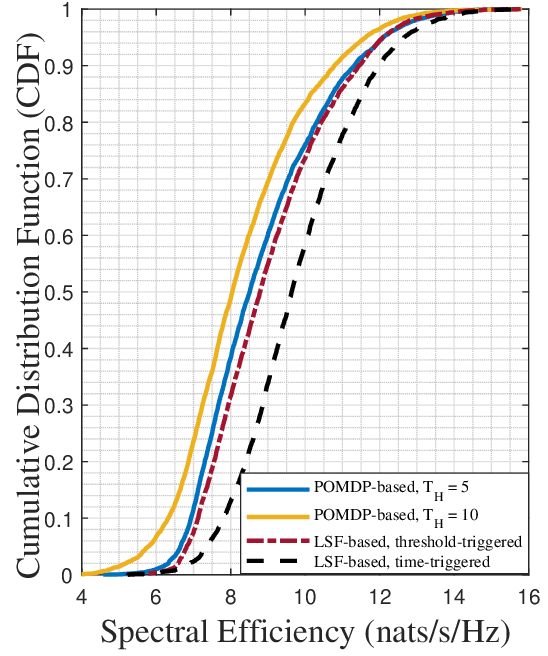}
			\vspace{-0.5em}
			\caption{Cumulative density function of spectral efficiency of typical user.}
			\label{fig:CDF_decHO}
		\end{subfigure}
		$\ $
		\begin{subfigure}[t]{0.31\textwidth}
			\centering			\includegraphics[width=1\textwidth]{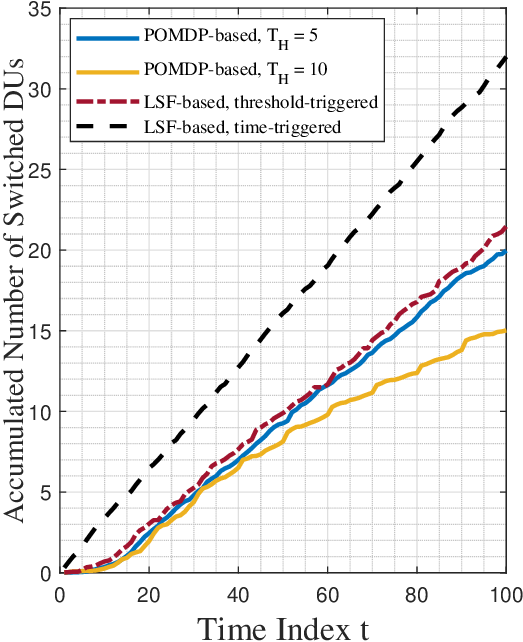}
			\vspace{-0.5em}
			\caption{Cumulative average number of switched APs.} 
			\label{fig:avgSwitchedDUs_cummulative_decHO}
		\end{subfigure}
		$\ $
		\begin{subfigure}[t]{0.315\textwidth}
			\centering			\includegraphics[width=1\textwidth]{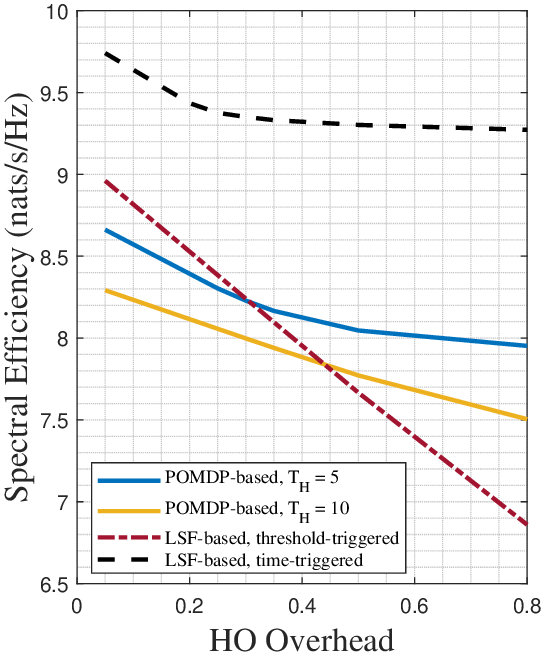}
			\vspace{-0.5em}
			\caption{spectral efficiency vs overhead of HOs.}
			\label{fig:rate_overhead}
		\end{subfigure}
		\caption{Performance using Algorithm~\ref{algorithm:DecreaseHO} at 
			$R_{\rm threshold} = 7$ nats/s/Hz.}
		\label{fig:ResultSet_decHO}
	\end{figure*}

	In Figs.~\ref{fig:ResultSet_decHO}(\subref{fig:CDF_decHO}) and~\ref{fig:ResultSet_decHO}(\subref{fig:avgSwitchedDUs_cummulative_decHO}), we show the CDF of the spectral efficiency and the accumulated number of switched APs, respectively. Our results show that using a time horizon $T_{\rm H} = 10$ decision cycles our proposed solution provides a number of HOs that is \nOfHOsTtrigLower\ lower compared to a time-triggered LSF-based HO policy and is \nOfHOsThreshtrig\ lower compared to a data rate threshold-triggered LSF-based HO policy. On the other hand, we observe an \CDFTtrigLower\ and \CDFThreshtrig\ decrease in the $10^{\rm th}$ percentile rate is observed compared to the two variants of LSF-based HO, respectively.

	In Fig.~\ref{fig:ResultSet_decHO}(\subref{fig:rate_overhead}), we plot the CDF of the spectral efficiency when each HO has a time overhead. The time overhead corresponds to the fraction of time spent in performing control operations, such as changing the serving set of the user and achieving synchronization between new serving APs. HOs could cause a disruption in the communication links which affects the performance and quality of service of the users. As the overhead of HOs can vary depending on many things, such as the HO algorithm, architecture of UC-mMIMO network, scenario of HO, etc, we measure the HO overhead as a fraction (unitless) of $\tau_c$, the conventional channel coherence interval.

	Our results show a possible advantage of minimizing the number of HOs using a POMDP framework, where the spectral efficiency of a POMDP-based HO scheme can outperform the LSF-based threshold triggered HO scheme at high HO overheads. What is interesting is that the POMDP-based HO scheme seems to be less affected by the increase in the overhead of the HOs compared to the LSF-based threshold triggered HO scheme. On the other hand, the LSF-based time-triggered HO scheme still provides a good performance in our studied scenario. However, the partial observability of the POMDP solution makes it more realistic than the LSF-based solution which requires knowing the LSF statistics to all APs. Moreover, the ability to employ an advanced framework, such as DRL, makes our solution more promising to manage HOs.

	\section{Conclusion}\label{section:Conclusion}
	We have formulated the problem of HO in user-centric cell-free MIMO networks as a POMDP. The model sets the combinations of the discrete version of the large-scale fading of the channels as the POMDP states, and the connection decisions between the user and the APs as the action space. We introduced an algorithm that controls the computational complexity as the network size increases through introducing the concept of POMDP sub-problems. We also introduced a method to decrease the number of HOs while maintaining a threshold for the quality of service. Our results show that our novel solution can control the number of HOs while maintaining a rate guarantee, where a~\nOfHOsTtrigLower-\nOfHOsThreshtrig\ reduction of the cumulative number of HOs is observed in networks with density of~$125$ APs per~${\rm km}^2$.

	For future studies, we have at least two possible directions. The first direction is to consider different quality of service metrics that include the handoff delays and the signaling overheads in the reward function. This will help us to study the effect of these metrics on the HO behavior. The second direction is to use DRL to manage HOs, which should be a scalable solution that can be deployed without a divide-and-conquer approach. The DRL approach will be built on the POMDP formulation and the theoretical formulation developed in this paper.

	\appendix
	\section{}
	\subsection{Proof for~\eqref{eq:rate_LB}: Lower Bound for Channel Capacity}\label{Appendix:rate_LB}
	The derivations in this appendix follow the same steps as those found in~\cite{CSIAgingzheng2020cell}  with some minor differences. The power of the desired signal can be derived as
	\allowdisplaybreaks{
		\begin{align}
			&\xi_{1,u}[n]
			 =
			\left| \mathbb{E}\left\{{\rm DS}_u[n]\right\} \right|^2
			\nonumber \\
			& =
			\Big|
			\rho_{u}[n - n_{\rm est}]
			\sum_{b \in \mathcal{C}_u} \sqrt{\eta_{bu}[n_{\rm est}]} 
			\mathbb{E}\left\{
			{\bf h}_{bu}^T[n_{\rm est}]
			{\bf \hat{h}}_{bu}^*[n_{\rm est}]
			\right\}
			\Big|^2
			\nonumber\\
			&=
			\Big|
			\rho_{u}[n - n_{\rm est}]
			\sum_{b \in \mathcal{C}_u} \sqrt{\eta_{bu}[n_{\rm est}]} 
			\nonumber\\
			& \quad \quad \quad \quad \quad \quad \quad
			\times
			\mathbb{E}\left\{
			(\mathrm{ {\bf e}}_{bu}^T[n_{\rm est}] + {\bf \hat{h}}_{bu}^T[n_{\rm est}])
			{\bf \hat{h}}_{bu}^*[n_{\rm est}]
			\right\}
			\Big|^2
			\nonumber\\
			&=
			M^2
			\rho_{u}^2[n - n_{\rm est}]
			\Big|
			\sum_{b \in \mathcal{C}_u} \sqrt{\eta_{bu}[n_{\rm est}]} 
			\psi_{bu}[n_{\rm est}]
			\Big|^2
			\nonumber \\
			&
			\stackrel{(a)}{=}
			M p^{({\rm d})}
			\rho_{u}^2[n - n_{\rm est}]
			\left|
			\sum_{b \in \mathcal{C}_u} \sqrt{
				\frac{\psi_{bu}[n_{\rm est}]}{|\mathcal{E}_b|}
			} 
			\right|^2
	\end{align}}
	where $\mathbb{E}\left\{
	\mathrm{ {\bf e}}_{bu}^T[n_{\rm est}]
	{\bf \hat{h}}_{bu}^*[n_{\rm est}]
	\right\} = 0$ because the two terms are uncorrelated, and $\mathbb{E}\left\{
	{\bf \hat{h}}_{bu}^T[n_{\rm est}]
	{\bf \hat{h}}_{bu}^*[n_{\rm est}]
	\right\} = M \psi_{bu}[n_{\rm est}]$, then $\mathbb{E}\left\{
	{\bf h}_{bu}^T[n_{\rm est}]
	{\bf \hat{h}}_{bu}^*[n_{\rm est}]
	\right\} = M \psi_{bu}[n_{\rm est}]$. The step in $(a)$ follows from~\eqref{eq:eta}.
	
	The beamformer uncertainty can be derived as
	\begin{align}\label{eq:BU_analysis}
		&\xi_{2,u}[n]
		= 
		\mathbb{E}\left\{\left|{\rm BU}_u[n]\right|^2\right\} 
		\nonumber \\
		& =
		\mathbb{E}\bigg\{\bigg|
		\rho_{u}[n - n_{\rm est}]\sum_{b \in \mathcal{C}_u} \sqrt{\eta_{bu}[n_{\rm est}]}  
		\nonumber\\[-5pt]
		&
		\quad \quad \quad
		\times
		\left(
		{\bf h}_{bu}^T[n_{\rm est}]
		{\bf \hat{h}}_{bu}^*[n_{\rm est}]
		-
		\mathbb{E}\left\{
		{\bf h}_{bu}^T[n_{\rm est}]
		{\bf \hat{h}}_{bu}^*[n_{\rm est}]
		\right\}
		\right)
		\bigg|^2\bigg\}
		\nonumber \\[-5pt]
		& \stackrel{(a)}{=}
		\rho_{u}^2[n - n_{\rm est}]
		\sum_{b \in \mathcal{C}_u} 
		\eta_{bu}[n_{\rm est}]  
		\nonumber\\[-5pt]
		&
		\ 
		\times
		\left(
		\mathbb{E}\left\{ \left| {\bf h}_{bu}^T[n_{\rm est}]
		{\bf \hat{h}}_{bu}^*[n_{\rm est}]
		\right|^2
		\right\}
		-
		\left|
		\mathbb{E}\left\{
		{\bf h}_{bu}^T[n_{\rm est}]
		{\bf \hat{h}}_{bu}^*[n_{\rm est}]
		\right\}\right|^2
		\right)
	\end{align}
	where $(a)$ follows because the variance of a sum of independent random variables equals the sum of the variances.
	
	The first term in~\eqref{eq:BU_analysis} can be calculated as
	\begin{align}\label{eq:BU_analysis2}
		&\mathbb{E}\left\{ \left| {\bf h}_{bu}^T[n_{\rm est}]
		{\bf \hat{h}}_{bu}^*[n_{\rm est}]
		\right|^2
		\right\}
		=
		\nonumber\\ 
		& \quad \quad \quad
		\mathbb{E}\left\{ \left| \mathrm{ {\bf e}}^T[n_{\rm est}]
		{\bf \hat{h}}_{bu}^*[n_{\rm est}]
		\right|^2
		\right\}
		+ \mathbb{E}\left\{ \left\|
		{\bf \hat{h}}_{bu}^*[n_{\rm est}]
		\right\|^4
		\right\}
		\nonumber \\
		&=
		\left(M^2  \psi_{bu}[n_{\rm est}] \left(\beta_{bu} - \psi_{bu}[n_{\rm est}] \right) \right)
		+
		2 M^2 (\psi_{bu}[n_{\rm est}])^2
	\end{align}
	The second term in~\eqref{eq:BU_analysis2} follows from the fact that for a Gaussian random variable (RV) $X\sim \mathcal{N}(0, \psi_{bu}[n_{\rm est}])$ the RV $Y = X^2$ has a chi-squared distribution. 
	Then,
	\begin{align}
		\xi_{2,u}[n]
		& =
		M^2 \rho_{u}^2[n - n_{\rm est}]
		\sum_{b \in \mathcal{C}_u} 
		\eta_{bu}[n_{\rm est}] 
		\beta_{bu}\psi_{bu}[n_{\rm est}]
	\end{align}
	The channel aging term can be calculated as
	{\allowdisplaybreaks\begin{align}
		&\xi_{3,u}[n]
		=
		\mathbb{E}\left\{\left| {\rm CA}_u[n] \right|^2\right\} 
		\nonumber \\[-5pt]
		& =
		\bar{\rho}_{u}^2[n - n_{\rm est}]
		\mathbb{E}\bigg\{ \Big|
		\sum_{b \in \mathcal{C}_u} \sqrt{\eta_{bu}[n_{\rm est}] \beta_{bu}} 
		{\bf v}_{bu}^T[n]
		{\bf \hat{h}}_{bu}^*[n_{\rm est}]
		\Big|^2\bigg\}
		\nonumber \\[-5pt]
		& =
		\bar{\rho}_{u}^2[n - n_{\rm est}]
		\left(
		\sum_{b \in \mathcal{C}_u} \eta_{bu}[n_{\rm est}] \beta_{bu} 
		\Xi_{1,bu}
		\right.
		\nonumber \\[-5pt]
		& \quad \quad
		+
		\left.
		\sum_{b \in \mathcal{C}_u}
		\sum_{b' \in \mathcal{C}_u, b' \ne b} 
		\sqrt{\eta_{bu}[n_{\rm est}] \beta_{bu}}
		\sqrt{\eta_{b'u}[n_{\rm est}] \beta_{b'u}}
		\Xi_{2,bb'u}
		\right)
	\end{align}
	with
	\begin{align}
		\Xi_{1, bu} &= \mathbb{E}\left\{ \left|
		{\bf v}_{bu}^T[n]
		{\bf \hat{h}}_{bu}^*[n_{\rm est}]
		\right|^2\right\}
		= M^2 \psi_{bu}[n_{\rm est}]
		\\
		\Xi_{2,bb'u} & =
		\mathbb{E}\left\{
		\left(
		{\bf v}_{bu}^T[n]
		{\bf \hat{h}}_{bu}^*[n_{\rm est}]\right)^*
		\left(
		{\bf v}_{b'u}^T[n]
		{\bf \hat{h}}_{b'u}^*[n_{\rm est}]\right)
		\right\} = 0
	\end{align}
	Then,
	\begin{align}
		\xi_{3,u}[n]
		=
		M^2 \bar{\rho}_{u}^2[n - n_{\rm est}]
		\sum_{b \in \mathcal{C}_u} \eta_{bu}[n_{\rm est}] \beta_{bu} 
		\psi_{bu}[n_{\rm est}]
	\end{align}}
	Using the definition of $\bar{\rho}_{u}[\cdot]$, we can combine $\xi_{2,u}$ and $\xi_{3,u}$ into a single expression as
	\begin{align}
		&\xi_{2,3,u}[n] =
		\xi_{2,u}[n] + \xi_{3,u}[n]
		\nonumber \\
		& \quad
		=
		M^2
		\sum_{b \in \mathcal{C}_u} \eta_{bu}[n_{\rm est}] \beta_{bu} 
		\psi_{bu}[n_{\rm est}]
		= 
		M
		\sum_{b \in \mathcal{C}_u} 
		\frac{p^{({\rm d})} \beta_{bu} }
		{
			|\mathcal{E}_b|
		}
	\end{align}
	The interference caused by the APs due to serving user $u'$ can be calculated as
	\begin{align}
		&\xi_{4,uu'}[n]
		=
		\mathbb{E}\left\{\left| {\rm MI}_{uu'}[n] \right|^2\right\} 
		\nonumber \\[-5pt]
		& = \mathbb{E}\bigg\{\bigg| \sum_{b' \in \mathcal{C}_{u'}} \sqrt{\eta_{b'u'}[n_{\rm est}]} {\bf h}_{b'u}^T[n] \hat{\bf h}_{b'u'}^{*}[n_{\rm est}] \bigg|^2\bigg\}
		\nonumber \\[-5pt]
		& =
		\sum_{b' \in \mathcal{C}_{u'}} \eta_{b'u'}[n_{\rm est}] \Xi_{3,b'uu'}
		\nonumber \\[-5pt]
		& \quad \quad
		+
		\sum_{b' \in \mathcal{C}_{u'}} \sum_{b'' \in \mathcal{C}_{u'}, b'' \ne b'} \sqrt{\eta_{b'u'}[n_{\rm est}]} \sqrt{\eta_{b''u'}[n_{\rm est}]} \Xi_{4,b'b''uu'}
	\end{align}
	where
	\begin{align}
		&\Xi_{3,b'uu'} = \mathbb{E}\left\{\left| {\bf h}_{b'u}^T[n] \hat{\bf h}_{b'u'}^{*}[n_{\rm est}] \right|^2\right\}
		\nonumber \\
		& \stackrel{(a)}{=}
		\rho_{u}^2[n - n_{\rm est}] \mathbb{E}\left\{\left| {\bf h}_{b'u}^T[n_{\rm est}] \hat{\bf h}_{b'u'}^{*}[n_{\rm est}] \right|^2\right\}
		\nonumber \\
		& \quad \quad \quad
		+
		\beta_{b'u}
		\bar{\rho}_{u}^2[n - n_{\rm est}]
		M^2 
		\psi_{b'u'}
		\nonumber \\
		& \stackrel{(b)}{=}
		\begin{cases}
			\begin{aligned}
			&
			\rho_{u}^2[n - n_{\rm est}] \left( M^2\beta_{b'u}\psi_{b'u'} + M^2 \psi_{b'u} \psi_{b'u'} \right)
			\nonumber \\[2pt]
			& \quad \quad \quad
			+
			\beta_{b'u}
			\bar{\rho}_{u}^2[n - n_{\rm est}]
			M^2 
			\psi_{b'u'},
			\end{aligned}
			& \text{if}\ u' \in \mathcal{U}_i
			\\[2pt]
			\begin{aligned}
			&
			\rho_{u}^2[n - n_{\rm est}] \left( M^2\beta_{b'u}\psi_{b'u'} \right)
			\nonumber \\[2pt]
			& \quad \quad \quad
			+
			\beta_{b'u}
			\bar{\rho}_{u}^2[n - n_{\rm est}]
			M^2 
			\psi_{b'u'},
			\end{aligned}
			& \text{if}\ u' \notin \mathcal{U}_i
		\end{cases}
		\nonumber \\
		& \stackrel{(c)}{=}
		\begin{cases}
			M^2\beta_{b'u}\psi_{b'u'} + M^2 \rho_{u}^2[n - n_{\rm est}] \psi_{b'u} \psi_{b'u'}
			,& \text{if}\ u' \in \mathcal{U}_i
			\\
			M^2 \beta_{b'u}\psi_{b'u'}
			,& \text{if}\ u' \notin \mathcal{U}_i
		\end{cases}
	\end{align}
	where $(a)$ is obtained using~\eqref{eq:channelEvolution_data}, $(b)$ follows from the expression of the estimated channel ${\bf \hat{h}}_{bu}$ in~\eqref{eq:est_chan} where user $u$ is assumed using the pilot $i$, i.e., $u \in \mathcal{U}_i$, and $(c)$ follows using the definition $\bar{\rho}_{u}[n - n_{\rm est}] = \sqrt{ 1 - |\rho_{u}[n - n_{\rm est}]|^2 }$. Also,
	\begin{align}	
		&\Xi_{4, b'b''uu'}
		=
		\mathbb{E}\left\{\left( {\bf h}_{b'u}^T[n] \hat{\bf h}_{b'u'}^{*}[n_{\rm est}] \right)^* \left( {\bf h}_{b''u}^T[n] \hat{\bf h}_{b''u'}^{*}[n_{\rm est}] \right)\right\}
		\nonumber \\[2pt]
		&
		=
		\resizebox{0.95\linewidth}{!}
		{$\displaystyle
		\rho_{u}^2[n - n_{\rm est}] \mathbb{E}\left\{ {\bf h}_{b'u}^T[n_{\rm est}] \hat{\bf h}_{b'u'}^{*}[n_{\rm est}] \right\}
		\mathbb{E}\left\{
		{\bf h}_{b''u}^T[n_{\rm est}] \hat{\bf h}_{b''u'}^{*}[n_{\rm est}] \right\}
		$}
		\nonumber\\[2pt]
		&=
		\begin{cases}
			\resizebox{0.7\linewidth}{!}
			{$\displaystyle
			\rho_{u}^2[n - n_{\rm est}] \left( M \sqrt{\psi_{b'u} \psi_{b'u'} } \right)
			\left( M \sqrt{\psi_{b''u} \psi_{b''u'} } \right)
			$}
			,& \text{if}\ u' \in \mathcal{U}_i
			\\
			0,& \text{if}\ u' \notin \mathcal{U}_i
		\end{cases}
	\end{align}
	
	Then,{\allowdisplaybreaks
		\begin{align}
			&\xi_{4,uu'}[n]
			=
			\nonumber\\
			&
			\begin{cases}
				\begin{aligned}
					&
					M^2 \bigg(\sum_{b' \in \mathcal{C}_{u'}} \eta_{b'u'}[n_{\rm est}]
					\beta_{b'u}\psi_{b'u'}
					+
					\rho_{u}^2[n - n_{\rm est}]
					\\
					&\quad \quad \quad 
					\times
					\bigg|\sum_{b' \in \mathcal{C}_{u'}} \sqrt{\eta_{b'u'}[n_{\rm est}]} \sqrt{\psi_{b'u} \psi_{b'u'} }
					\bigg|^2
					\bigg)
				\end{aligned}
				,& \text{if}\ u' \in \mathcal{U}_i
				\\
				M^2 \sum_{b' \in \mathcal{C}_{u'}} \eta_{b'u'}[n_{\rm est}]
				\beta_{b'u}\psi_{b'u'}
				,& \text{if}\ u' \notin \mathcal{U}_i
			\end{cases}
			\nonumber\\
			&=
			\begin{cases}
				\begin{aligned}
					&
					M \bigg(\sum_{b' \in \mathcal{C}_{u'}} \frac{p^{({\rm d})}}
					{
						|\mathcal{E}_{b'}|
					}
					\beta_{b'u}
					\\
					&\quad \quad  
					+
					\rho_{u}^2[n - n_{\rm est}]
					\bigg|\sum_{b' \in \mathcal{C}_{u'}} \sqrt{\frac{p^{({\rm d})} \psi_{b'u} }
						{
							|\mathcal{E}_{b'}|
						}
					}
					\bigg|^2
					\bigg)
				\end{aligned}
				,& \text{if}\ u' \in \mathcal{U}_i
				\\
				\displaystyle
				M \sum_{b' \in \mathcal{C}_{u'}} 
				\frac{p^{({\rm d})}}
				{
					|\mathcal{E}_{b'}|
				}
				\beta_{b'u}
				,& \text{if}\ u' \notin \mathcal{U}_i
			\end{cases}
	\end{align}}
	Then, the final formula for the rate in~\eqref{eq:rate_LB} is obtained by representing the summation terms using vector multiplication.

	\vspace{-1em}
	\subsection{Spatial Correlation of Large-scale Fading}
	\label{sec:SpatialCorrelationLSF}
	The LSF between AP $b$ and user $u$ is defined as
	$\beta_{bu} \triangleq {\rm PL} \times 10^{\frac{\sigma_\mathrm{sh|dB} \bar{\kappa}_{bu}}{10}}$, 
	where ${\rm PL}$ accounts for the effect of the path loss, while the remaining term accounts for the shadowing with $\sigma_\mathrm{sh|dB}$ being the standard deviation of the shadowing in dB scale and $\bar{\kappa}_{bu}$ is defined next.
	
	We use a two-component shadowing model that is validated in~\cite{CorrelatedShadowing4357088, CorrelatedShadowing104090}. The model accounts for the correlation in shadowing obtained from having common obstacles between the transmitters and the receivers when they are in close vicinity. 
	This model is represented through $\bar{\kappa}_{bu}$~as
	\begin{align}
		\bar{\kappa}_{bu} = \sqrt{\iota} \kappa_{1,b} + \sqrt{1 - \iota} \kappa_{2,u},
	\end{align}
	where $\kappa_{1,b},\kappa_{2,u} \sim \mathcal{N}\left(0, 1\right)$, 
	and $0 \le \iota \le 1$. The term $\kappa_{1,b}$ models the contribution of the obstacles in the vicinity of AP $b$ to the shadowing experienced by all users, while $\kappa_{2,u}$ models those in the vicinity of user $u$ and affects the shadowing to all the APs. The covariance between $\kappa_{1,b}$ and $\kappa_{1,b'}$, and between $\kappa_{2,u}$ and $\kappa_{2,u'}$ are, respectively, given by~\cite{cellFreeUserCentricPower8901451}
	\begin{align}\label{eq:correl_shad}
		C_{\kappa_{1,b},\kappa_{1,b'}}
		&\triangleq
		\mathbb{E}\left\{\kappa_{1,b} \kappa_{1,b'} \right\} = 2^{-\frac{d_{1,b b'}}{ d_\mathrm{decorr}}}\ \text{and} 
		\nonumber\\
		C_{ \kappa_{2,u},\kappa_{2,u'} }
		&\triangleq
		\mathbb{E}\left\{\kappa_{2,u} \kappa_{2,u'} \right\} = 2^{-\frac{d_{2,u u'}}{ d_\mathrm{decorr}}}.
	\end{align}
	The parameter $d_{1,b b'}$ is the distance between APs $b$ and $b'$, and $d_{2,u u'}$ is the distance between users $u$ and $u'$. The decorrelation distance $d_\mathrm{decorr}$ has typical values between $20$ and $200$ ${\rm m}$, and it corresponds to the distance at which the correlation drops to $50\%$, where smaller values correspond to an environment with a lower degree of stationarity.
	
	Using equation~\eqref{eq:correl_shad}, we redefine $\mathbb{E}\left\{\kappa_{2,u} \kappa_{2,u'} \right\}$ for a moving user $u$ as
	\begin{align}\label{eq:correl_shad_2}
		C_{ \kappa_{2,u}^{(t)}, \kappa_{2,u}^{(t-1)} } = \mathbb{E}\left\{\kappa_{2,u}^{(t)} \kappa_{2,u}^{(t-1)} \right\} =
		2^{-\frac{\mathtt{v}_u \bar{\Delta} }{ d_\mathrm{decorr}}},\ t>0
	\end{align}
	where $\bar{\Delta}$ is the time difference between two consecutive time indices, e.g., between $t$ and $(t-1)$, and $\mathtt{v}_u$ is the velocity of user $u$. The intuition from~\eqref{eq:correl_shad_2} is that during the movement of any user, the shadowing experienced by the user before and after the movement will be correlated due to the same objects found in the area. This allows us to construct a realistic model to evolve the shadowing during the movement of the user. Later on, we will show how this model have an important usage for our HO problem.
	
	
	Based on this, we define the shadowing experienced by the user at $t$ as
	\begin{align}\label{eq:shad_model}
		\bar{\kappa}_{bu}^{(t)} =
		\begin{cases}
			\sqrt{\iota} \kappa_{1,b} + \sqrt{1 - \iota} \kappa, & \text{if}\ t = 0\\ 
			\begin{aligned}
			&\sqrt{\iota} \kappa_{1,b} + \sqrt{1 - \iota} \kappa_{2,u}^{(t)}, 
			\\
			& \quad
			\text{with}\  C_{ \kappa_{2,u}^{(t)}, \kappa_{2,u}^{(t-1)} }\ \text{defined in}~\eqref{eq:correl_shad_2},
			\end{aligned}
			& \text{if}\ t > 0
		\end{cases}	
	\end{align}
	where $\kappa \sim \mathcal{N}(0, 1)$.
	
	For the path loss, we define ${\rm PL}$ using the COST231 Walfisch-Ikegami model~\cite{Walfisch14401} as
	\begin{align}\label{eq:PL}
		{\rm PL}^{(t)} =
		\left(\sqrt{\left(\bar{d}_{bu}^{(t)}\right)^2 + d_{\rm h}^2}\right)^{- \alpha_{\rm pl} } {d_0}^{\alpha_{\rm pl} }
		,
	\end{align}
	where $\bar{d}_{bu}^{(t)}$ is the geographical distance between AP $b$ and user $u$ at time index $t$ in the \emph{xy}-plane. 
	The term $d_{\rm h}$ is the fixed difference in the heights between the APs and the user and is used to impose an exclusion region between the APs and the users while the user is moving. Furthermore, for an operating frequency of $1.8$~GHz, the values of the path loss exponent $\alpha_{\rm pl} = 3.8$ and the path loss reference distance $d_0 = 1.1$~meters.
	
	
	\noindent Finally, the temporally correlated LSF between AP $b$ and user $u$ at $t$ can be summarized as:
	\begin{align}\label{eq:largeScaleFading}
		\beta_{bu}^{(t)} =
		{\rm PL}^{(t)} \times 10^{\frac{\sigma_\mathrm{sh|dB} \bar{\kappa}_{bu}^{(t)}}{10}}.
	\end{align}
	where $\bar{\kappa}_{bu}^{(t)}$ and ${\rm PL}^{(t)}$ are defined in~\eqref{eq:shad_model} and~\eqref{eq:PL}, respectively.

	\footnotesize
	\bibliography{Mob_HO_References}
	\bibliographystyle{ieeetr}

\end{document}